\documentclass[12pt,preprint]{aastex}

\usepackage{amsmath,natbib}
\usepackage{rotating}
\usepackage{graphicx}

\newcommand{\bvec}[1]{\mbox{\boldmath $#1$}}

\renewcommand{\vec}[1]{{\mathbf #1}}

\def\cov{ {\bf C}}
\def\csrc{ {\bf C}^{src}}
\def\cscan{ {\bf C}^{scan}}
\def\data{ \bvec{d}}

\def\nrun{ \frac{d n_s}{d\log k}}

\def\eg{{\it e.g.,\,}}

\def\ujy{\, {\rm \mu Jy}}

\def\ie{{\it i.e.,\,}}

\def\spose#1{\hbox to 0pt{#1\hss}}
\def\lta{\mathrel{\spose{\lower 3pt\hbox{$\mathchar"218$}}
     \raise 2.0pt\hbox{$\mathchar"13C$}}}
\def\gta{\mathrel{\spose{\lower 3pt\hbox{$\mathchar"218$}}
     \raise 2.0pt\hbox{$\mathchar"13E$}}}

\def\uvp{$(u,v)$-plane}

\newcommand{\jysq}{\rm{Jy}^2/\rm{sr}}

\newcommand{\KS}{KSSZ}
\newcommand{\SPH}{SPH}

\slugcomment{submitted to ApJ}

\begin{document}

\title{Cosmological Results from Five Years of 30 GHz CMB Intensity
  Measurements with the Cosmic Background Imager}

\author{J.L. Sievers\altaffilmark{1}, 
 B.S. Mason\altaffilmark{2},
 L. Weintraub\altaffilmark{3}, 
 C. Achermann\altaffilmark{4},
 P. Altamirano\altaffilmark{4},
 J.R. Bond\altaffilmark{1},
 L. Bronfman\altaffilmark{4},
 R. Bustos\altaffilmark{13}, 
 C. Contaldi\altaffilmark{7},
 C. Dickinson\altaffilmark{3,8},
 M.E. Jones\altaffilmark{9},
 J. May\altaffilmark{4}, 
 S.T. Myers\altaffilmark{11},
 N. Oyarce\altaffilmark{4},
 S. Padin\altaffilmark{6}, 
 T.J. Pearson\altaffilmark{3}, 
 M. Pospieszalski\altaffilmark{12},
 A.C.S. Readhead\altaffilmark{3},
 R. Reeves\altaffilmark{13},
 M. C. Shepherd\altaffilmark{3}, 
 A. C. Taylor\altaffilmark{9},
 S. Torres\altaffilmark{13}
}

\altaffiltext{1}{Canadian Institute for Theoretical
Astrophysics, University of Toronto, ON M5S 3H8, Canada} 
\altaffiltext{2}{National Radio Astronomy Observatory, 520 Edgemont Road, Charlottesville, VA 22903}
\altaffiltext{3}{Owens Valley Radio Observatory, California Institute of
Technology, Pasadena, CA}
\altaffiltext{4}{Departamento de Astronom{\'\i}a, Universidad de Chile,  Santiago, Chile}
\altaffiltext{6}{Kavli Institute for Cosmological Physics, Department of Astronomy and Astrophysics, 
University of Chicago, Chicago, IL 60637}
\altaffiltext{7}{Department of Physics, Imperial College, London, UK}
\altaffiltext{8}{Infrared Processing \& Analysis Center, California Institute of Technology, M/S 220-6, 1200 E. California Blvd., Pasadena, CA 91125}
\altaffiltext{9}{Astrophysics, Oxford University, Keble Road, Oxford OX1 3RH, UK} 
\altaffiltext{10}{U.C. Berkeley Astronomy}
\altaffiltext{11}{National Radio Astronomy Observatory, Socorro, NM 87801} 
\altaffiltext{12}{NRAO New Technology Center, Charlottesville VA 22903}
\altaffiltext{13}{Departamento de Ingenier{\'\i}a El{\'e}ctrica, Universidad de Concepci{\'o}n, Concepci{\'o}n, Chile}

\begin{abstract}

  We present final results on the angular power spectrum of total
  intensity anisotropies in the Microwave Background from the Cosmic
  Background Imager (CBI). Our analysis includes all primordial
  anisotropy data collected between January 2000 and April 2005, and
  benefits significantly from an improved maximum likelihood analysis
  pipeline.  It also includes results from a 30 GHz foreground survey
  conducted with the Green Bank Telescope (GBT) which places
  significant constraints on the possible contamination due to
  foreground point sources.  We improve on previous CBI results by
  about a factor of two in the damping tail.  These data confirm, at
  $\sim 3\sigma$, the existence of an excess of power over intrinsic
  CMB anisotropy on small angular scales ($\ell > 1800$).  Using the
  GBT survey, we find currently known radio source populations are not
  capable of generating the power; a new population of faint sources
  with steeply rising spectral indices would be required to explain
  the excess with sources.  Extensive testing does not reveal any
  instrumental effect capable of giving rise to the observed excess.
  We also present a full cosmological parameter analysis of the new
  CBI power spectrum, the WMAP 5-year data, and the latest ACBAR data,
  self-consistently including in the analysis the foreground signal
  from the Sunyaev-Zel'dovich Effect (SZE) from galaxy clusters.  With
  CBI alone, the full parameter analysis finds the excess is
  $1.6\sigma$ above the level expected for a $\sigma_8=0.8$ universe.
  We fit two different SZ templates to the power spectrum and find
  they give markedly different inferred $\sigma_8$ values suggesting
  more theoretical work is required. We find the addition of
  high-$\ell$ CMB data substantially improves constraints on cosmic
  string contributions to the TT power spectrum as well as the running
  of the scalar spectral index $n_{run}$, but that $n_{run}$ is quite
  sensitive to the level of the SZE. We also present forecasts for
  what other experiments should see at different frequencies and
  angular resolutions given the excess power observed by CBI. We find
  that the reported high $\ell$ bandpowers from current high
  resolution CMB bolometer experiments are consistent with each other
  and CBI if the excess power is due to the SZE at the CBI-level of
  $2.5\pm 1$ times the $\sigma_8=0.8$ standard SZ template. This is
  not the case if the CBI excess source has a flat frequency
  dependence in thermodynamic temperature.

\end{abstract}



\noindent

\section{Introduction}

Measurements of Cosmic Microwave Background (CMB) anisotropies in
total intensity
\citep{deBernardis:2000gy,Hanany:2000qf,Halverson:2001yy,paper3,Dickinson:2004yr,Nolta08}
and polarization \citep{Kovac:2002fg,paper8,paper9,Montroy:2005,quad07,quad08,Nolta08} over the
past decade-- together with a range of other cosmological measurements
\citep{Riess:1998,Perlmutter:1998np,Freedman:2000cf}-- have provided
striking confirmation of the inflationary structure formation
paradigm. Key elements of this picture which have been confirmed are:
that the universe is spatially flat; that anisotropies in the
microwave background formed via simple causal processes from an
approximately scale-invariant spectrum of (probably adiabatic)
primordial inhomogeneities; and that the gravitational instability
picture of subsequent structure formation from the collapse of
baryons and dark matter in an expanding 
metric. In spite of what must, on the whole, be described as stunning
experimental confirmation of the inflationary predictions, several
surprises emerged, and tensions and ambiguities in our understanding
of the data remain.  Chief amongst these are the necessity for a
(presently dynamically dominant) dark energy component, some anomalies
in the large-angle WMAP data, and a $\sim 3
\sigma$ excess of power on small angular scales
\citep{paper2,paper7,kuo02,dawson06,kuo07, acbar08}. A number of secondary
anisotropies--- predominantly the Sunyaev-Zel'dovich Effect (SZE) from
galaxy clusters--- are expected to contribute on small angular scales,
thus providing a view of the more recent evolution of large-scale
structure through the CMB.

The Cosmic Background Imager (CBI) is a 30 GHz interferometer that
measures CMB anisotropies from $\ell \sim 400$ to $\ell \sim
3000$. The instrument has been described in detail in
\citet{Padin02}. From January 2000 through November of 2001 the CBI
surveyed $\sim 98 {\rm deg^2}$ of sky. Results from this work were
presented in \citet{paper1} (hereafter Paper 1), \citet{paper2} (Paper
2), \citet{paper3} (Paper 3), and \citet{paper7} (Paper 7).  Following
this campaign the instrument was upgraded to focus primarily on CMB
polarization observations, which were conducted from September 2002
through April 2005. The fields observed in the polarization
campaign encompassed $\sim 115 {\rm deg^2}$ in all, and partially
overlapped with the fields observed in the total-intensity campaign.
The total area covered in the combined datasets is 143${\rm deg^2}$. 
While roughly half of the CBI baselines were cross-polarized for
these observations (appropriate to measure polarization), the other
half were co-polar and thus improve results on the TT power spectrum.
Results from the polarization campaign are presented in \citet{paper8}
(Paper 8) and \citet{paper9} (Paper 9).  In this paper we combine all
of these data to present the final TT power spectrum from five years
of CBI measurements. We also examine the implications of the observed
high-$\ell$ signal in the context of a full cosmological parameter
analysis including a treatment of uncertainties in the SZ foreground
models.

A key limiting factor in interpreting the high-$\ell$ excess measured
by CBI has been uncertainties associated with the extragalactic point
source correction. Extragalactic sources reduced the sensitivity of
the CBI at high-$\ell$ in two ways: by requiring a substantial amount
of data to be ``thrown out'' owing to possible contamination from known
sources, and through the uncertainty in the power spectrum of fainter
sources extrapolated from number counts. In order to address the second
issue, we conducted a 30 GHz survey with the GBT and the OVRO 40-m
telescope, covering a total of 3562 NVSS \citep{condon} sources in the
CBI fields; this survey is 
described in the companion paper \citet{paper11}.  This survey has
allowed us to place much tighter constraints on the point source
foreground contribution to our power spectrum, which we have
quantified and included in this analysis.  Because of the potential of
source variability, we do not attempt to reclaim the sky area under
NVSS sources that are not detected by the GBT.

The structure of the paper is as follows. In \S~\ref{sec:obs} we
briefly summarize the data.  \S~\ref{sec:analys} provides a
description of improvements to our analysis algorithms, describes
tests performed on the data, and summarizes our knowledge of
foregrounds (chiefly discrete sources) in the CBI fields. In
\S~\ref{sec:results} we present the CBI power spectrum, discuss the
significance and characteristics of the high-$\ell$ excess signal, and
give constraints on cosmological parameters from this.
\S~\ref{sec:summary} summarizes our findings.  We describe our
spectrum-fitting procedure in Appendix \ref{sec:maxlike}.

\section{Observations}
\label{sec:obs}

For this analysis we use data collected on both the CBI total
intensity fields from \citet{paper7} and the intensity data from the CBI
polarization fields described in \citet{paper9}.  The fields, observing
strategies, data reduction, and calibration are described in detail in Papers 2, 3, 7,
8, and 9; here we simply summarize the contents of these papers.

The CBI total intensity and polarization fields were spaced by $\approx 6$ hours
in Right Ascension (centered near 02h, 08h, 14h, and 20h) and are near
the celestial equator. The polarization and total intensity fields
overlap but are not identical.  To test for systematics and obtain
higher signal-to-noise ratio at high-$\ell$, some very deep integrations on
individual pointings were performed in each campaign. During the total
intensity campaign, all integration time at 8h was concentrated in a
single pointing; the 14h and 20h mosaic fields also contained one
pointing each on which significantly more integration time was
acquired. During the polarization campaign, observations at 20h
concentrated on a single 1$\times$6 strip instead of 6$\times$6 mosaics.  These
fields are summarized in Table~\ref{tbl:fields}. The deep 14h and 20h
fields, while within the total intensity mosaics of these regions and
collected and analyzed identically, are listed separately in this
table since for some tests we analyze only the deep data.

The largest source of systematic errors in the CBI data is a ground spillover signal,
which is strongest on the short baselines. We find this signal is stable over
timescales of at least 20 minutes.  For the total intensity
observations, contaminating ground signal was removed by differencing
pairs of individual pointings separated by 8 minutes in Right
Ascension. For the polarization mosaics 6 successive pointings were
observed in succession and a single common mode removed from matched
visibilities.

All data are calibrated with respect to the 5-year WMAP Ka Jupiter brightness
temperature\footnote{This is the 33 GHz Rayleigh-Jeans
brightness temperature of the planet minus the Rayleigh-Jeans
brightness temperature of the CMB at the same frequency.}
$T_{Jupiter}=146.6 \pm 0.75 \, {\rm K}$ \citep{Hill08}.  We adopt a
1\% in amplitude calibration uncertainty.  This compares with our
previous Jupiter estimate of $T_{Jupiter}=147.3 \pm 1.8 \rm{K}$ and
calibration uncertainty of 1.3\% \citep{paper7}.  

\section{Data Analysis}
\label{sec:analys}

\subsection{Improvements to the Analysis Pipeline}
\label{subsec:pipeline}

We use a maximum-likelihood algorithm to calculate the power spectrum.
It is based on the framework described in \citet{Myers03} (hereafter Paper~4),
but includes modifications to deal with the heterogeneous nature of the combined
CBI dataset.  The pipeline consists of two stages:  the first stage
compresses hundreds of thousands of visibilities into $\sim 10^4$ 
``gridded estimators'', and calculates noise, CMB signal, and point source
covariances for those estimators; the second stage calculates the
maximum-likelihood spectrum from the estimators, as well as ancillary
data products.

\subsubsection{Compressing the Data}
\label{sec:gridder}

As is the case for any compact array, the visibilities in the CBI data
have highly correlated signals.  To compress the data, we grid the
visibilities onto a set of regularly spaced estimators in the \uvp,
using the program CBIGRIDR (described in Paper~4).  This
compresses the few hundred thousand visibilities in a typical CBI
mosaic into a few thousand estimators with essentially no loss of
information, under the assumption of Gaussian noise.  One of the
outputs of CBIGRIDR is the matrix ${\bf R}$ which  
maps the (Fourier-plane) sky map into estimators, 
\begin{equation}\label{eq:deltaR}
  \bvec{\Delta} = {\bf R}\,\bvec{t} + \bvec{n},
\end{equation}
where $\bvec{t}$ is the Fourier transform of the true sky image, 
$\bvec{\Delta}$ is the gridded estimator vector, and
$\bvec{n}$ is the noise component of the gridded estimator (Equation~23 of Paper~4). 

The covariance of the gridded data vector $\bvec{\Delta}$ has signal and noise components
\begin{equation}\label{eq:ccovar}
  {\bf C} = \langle \bvec{\Delta}\,\bvec{\Delta}^{\dag} \rangle
          =  {\bf C}^{N} + {\bf C}^{T} + {\bf C}^{\rm scan} + {\bf C}^{\rm src} + {\bf C}^{\rm res}
\end{equation}
with contributions from noise, CMB, scan-dependent systematic errors, discrete point 
sources, and residual foregrounds (including a ``field'' of weak, confused point sources 
too faint for the CBI to detect individually) respectively.  The $\dag$ operator denotes the Hermitian conjugate 
(complex conjugate of the matrix or vector transpose).
The signal covariance ${\bf C}^{T}$ due
to the CMB temperature anisotropy signal is
\begin{equation}
{\bf C}^{T} = {\bf R}\,{\bf T}\,{\bf R}^\dag ,
\end{equation}
where 
\begin{equation}
{\bf T} \equiv \langle \bvec{t}\,\bvec{t}^\dag \rangle.
\end{equation}
For a Gaussian random CMB temperature field, we
expect ${\bf T}$ to be diagonal if $\bf{t}$ is represented in
spherical harmonic space, as a set of $a_{\ell,m}$'s, and thus
\begin{equation}
\langle t_{lm}t_{l'm'}\rangle = C_\ell \, \delta(\ell, \ell') \delta(m,m')
\end{equation}
encodes the CMB power spectrum $C_\ell$.  This remains true in the
small-angle approximation where 
\begin{equation}
  T_{ll'} =  \langle \vec{t}(\bvec{u}_{l})\,\bvec{t}^\ast(\bvec{u}_{l'}) \rangle
     = C_\ell \, \delta(\bvec{u}_{l}-\bvec{u}_{l'})
\end{equation}
 when $\bf{t}$ is represented in Fourier space and $\bvec{u}$ is
 the 2-D wavevector, with $\ell=2\pi |\bvec{u}|$ 
\citep{White99}.  Since $R$ contains all 
the information about the effects of the instrument (such as the
primary beam) and the scan strategy on the data, to combine
heterogeneous datasets, we simply grid them separately and then
combine their ${\bf R}$'s.  

In practice, the real and imaginary parts of the complex estimators
$\bvec{\Delta}$ are unpacked into a real estimator vector $\bvec{d}$, which
is then used for the covariance analysis and calculation of the
angular power spectrum.

\subsubsection{Calculation of the Power Spectrum}
\label{sec:mpilikely}

The log-likelihood of Gaussian data given the model for its signal and
noise given in Equation \ref{eq:ccovar} is 
\begin{equation}
\log \left ( \mathcal{L}\right ) = -\frac{1}{2}\bvec{\Delta}^{\dag} {\bf
  C}^{-1} \bvec{\Delta} - \frac{1}{2} \log \left ( | {\bf C} | \right ).
\end{equation}
The maximum-likelihood solution is the set of parameters $q_B$
which define the theory covariance $\cov_{T}\left ( q_B \right )$
such that the likelihood is maximized.  We restrict ourselves to
models of $ \cov_T$ of the form $\cov_T=\sum q_B \cov_B$.  One model
of this form is if the $\cov_B$ are bins in $\ell$, in which case the
$q_B$ are the binned power spectrum.

We calculate the maximum-likelihood CMB power spectrum from the
estimators using MPILIKELY, an MPI implementation of an algorithm
described in \citet{sieversthesis} and in Appendix
\ref{sec:maxlike}.  It requires a single matrix inversion, with no
additional $n^3$ operations to calculate the gradient and approximate
curvature of the likelihood at a trial spectrum.  We operate on real
data, as the correlations for the real and imaginary components of
visibilities can be different (see \citet{Myers03}).

The one algorithmic change we have made in MPILIKELY with the
potential to affect the output power spectrum is in the numerical
treatment of the projection of ground spillover and point sources with
known positions.  To project sources with known positions, but unknown
fluxes, we form the matrix $\csrc=\sum s_i s_i^T$, where $s_i$
is the expected signal from the $i^{th}$ source, and in the past have
added $\beta \csrc$ to the noise, for large $\beta$.  Similarly, for data taken
in strips with common ground (\ie that from \citet{paper9}), we calculate the
matrix $\cscan$ expected from the ground, and have also added $\gamma
\cscan$ with a large $\gamma$ to
the covariance.  By calculating the spectrum expected from the CMB
(see \ref{sec:spec_expected}), we find that the spectrum calculated
using large but finite values for both $\beta$ and $\gamma$ leads to a
recovered CMB power spectrum that is biased slightly low in the damping tail
due to numerical artifacts in matrix inversion - see Figure
\ref{fig:Bias}.  We now take the analytic limit as $\beta \rightarrow
\infty $ (see Appendix \ref{sec:sourceproj}) in MPILIKELY, and explicitly subtract
the ground signal as described in Section \ref{sec:obs}, with CBIGRIDR
accounting for the correlations induced by the subtraction.  As can be
seen in Figure \ref{fig:Bias}, the ground subtraction and analytic
source projection recovers an unbiased CMB power spectrum with
slightly reduced errors.

\subsection{Point Sources}
\label{sec:pointsources}

Point sources are the largest astrophysical foreground in the CBI
data, and are especially important at high-$\ell$.  All sources with
positions that are known reliably from low frequency radio
observations are removed from our power spectrum analysis; sources
below this threshold require a statistical correction to the power
spectrum.  An accompanying paper describes a campaign of GBT and OVRO
31 GHz measurements of $2,125$ NVSS sources in the CBI fields
\citep{paper11}.  As a result of this campaign we are able to much
more accurately determine this correction and characterize its
uncertainty.

The NRAO VLA Sky Survey \citep[NVSS][]{condon} is a $1.4$ GHz survey
of the northern sky, taken to be complete down to $S_{1.4} = 3.4 \,
{\rm mJy}$ (although its nominal detection limit is $2.5 \, {\rm
mJy}$).  All sources with integrated flux densities above $3.4$ mJy
in the NVSS catalog which lie within one degree of a CBI field are
projected.  Source projection works well as long as the relative
responses to the source are known for all data.  This is true if
either the source flux is constant for all observations, \textit{or}
all the data are taken in similar modes (same beam, observing strategy
etc.).  Because 30 GHz sources can be quite variable
\citep[e.g.][]{Cleary05} and the UV coverage and scan strategy changed
between the two CBI data sets, we project all sources separately from the
scan- and differenced-data.  In other words, we project two vectors
for each source, one corresponding to the source in the
differenced-data, and one in the scan-data.  

Sources fainter than $3.4$ mJy at $1.4$ GHz must be accounted for
statistically.  Although their sky density is statistically well
characterized down to $\ujy$ levels \citep[e.g.][]{hopkins03} from deep
observations in a variety of fields substantially smaller than the CBI
fields, these sources are the major systematic uncertainty in the
power at high-$\ell$ owing to the need for a spectral extrapolation
from $1.4$ GHz, where their counts are well known and where they are
{\it selected} for inclusion in the statistical term, to 31 GHz.  Our
analysis of GBT and OVRO 31 GHz flux density measurements in
comparison with the NVSS $1.4$ GHz values yields an average flux
density ratio $f \equiv S_{31}/S_{1.4} = 0.111 \pm 0.003$; using the
low frequency counts and the full distribution of 
$f$ (thus including its intrinsic width as constrained by our point
source dataset) we determine a point source correction $0.046 \pm
0.018 {\rm Jy^2/Sr}$ at 31 GHz.  The probability of an extreme source event
giving rise to the high-$\ell$ power is also low -- the highest
level of power in any of more than 200 realizations of sources
consistent with our data is less than $0.1 \jysq$ (see \citet{paper11}
for the full non-Gaussian distribution).  Our new source power is
nearly a factor of two lower than the earlier statistical correction
used in earlier CBI analyses, $0.08 \pm 0.04 \jysq$ \citep{paper2}, although
consistent within uncertainties. The earlier correction was a
conservative estimate based on a spectral index distribution which was
biased against steep-spectrum sources.  The current determination is
based on simulations proceeding from mock CBI observations of the
given population of residual sources through the full power spectrum
pipeline. Amongst other effects this includes the effects of Poisson
uncertainty in the faint source population. The full distribution of
the source correction is taken into account when we estimate
cosmological parameters.  Further details on the point source
observations and simulations are in \citet{paper11}.

Although our GBT and OVRO data provide strong constraints on the
spectral properties of the mJy-level extragalactic sources that
dominate the CBI statistical correction, it is possible that at
fainter flux densities the sources have different characteristics at
30 GHz.  The majority of the fainter $1.4$ GHz sources are expected to
have steep radio spectra \citep{richards00,condon}, however, were there
to be a substantial enhancement of sources with strongly inverted
spectra between $1.4$ and $31$ GHz they could contribute appreciably
to the excess. \citet{paper11} considers this scenario in detail,
finding that for moderately inverted spectra ($\alpha \sim 0.2$) they
would need to constitute 40\% of the sub-mJy population in order to
fully explain the CBI excess. Were the sources to have strongly
inverted spectra ($\alpha \sim 0.8$), 2\% of the population would be
required. In contrast, in the GBT+OVRO surveys, the most steeply inverted
spectrum source had $\alpha = 0.49$ and $<0.1\%$ of sources had
$\alpha > 0.3$.

\subsection{Data Tests}
\label{subsec:splits}

The data presented here have been discussed previously in
\citet{paper2}, \citet{paper3}, \citet{paper7}, \citet{paper8}, and
\citet{paper9}, where extensive data integrity tests were described.
A number of further tests have been carried out on the data. Key results
from this exercise are as follows:
\begin{enumerate}
\item {\bf Dish Pointing Errors:} By analyzing beam maps on bright sources we
have determined the individual antenna primary-beam pointing errors to be $\sim 3.5'$ RMS in a single
direction. We have simulated the impact of this, folding in the correlations induced
by deck rotations throughout our observing strategy along with our typical
$15''$ RMS pointing errors, and find that the resulting bias in the power spectrum is
$< 9 \, {\rm \mu K^2}$ in the highest $\ell$ bin. 
\item {\bf Spectral Index of Projected Sources:} We have introduced systematic
errors of $\delta \alpha=\pm 1$ in the spectral index used to project sources out of the
data and find that this has an effect $< 6 \, {\rm \mu K^2}$ in the highest $\ell$ bin.
\item {\bf Primary Beam:} To test the sensitivity of our power
spectrum to the precise beam shape used in the analysis, we ran simulations
of CBI observations using our best-fit physical model of the CBI beam 
\citep[described in][]{paper3}, but analyzed them with the best-fit Gaussian beam.
This results in $< 10 \, {\rm \mu K^2}$ error at any $\ell$.
\item {\bf Noise Fitting:} The way that we calculate the thermal noise
  for individual visibilities in our dataset results in a known $\sim
  2-6\%$ 
 underestimate of the thermal noise variance.  This underestimate depends on the
  observing strategy that was used. Simulations and analytic
  calculations of this effect have been presented in \citet{paper2}
  and \citet{sieversthesis}. To further constrain the thermal noise
  power spectrum, which can be a limiting factor especially at the
  highest $\ell$ CBI measures, we have implemented a noise estimator
  after the gridding step in the analysis pipeline. This estimator
  splits the 10 CBI frequency channels into two groups ({\it e.g.},
  low-frequency/high-frequency, or even/odd channels) and fits for a
  thermal noise multiplier and the CMB power spectrum
  simultaneously. For the polarization observations this exercise
  yields noise spectrum multipliers of $ 1.0195 +/- 0.009$ (low/high)
  and $1.015 +/- 0.009$ (even/odd), in comparison with our previous
  best estimate of $1.0175$. For the total intensity observations we
  find $1.071+/0.008$ and $1.071+/-0.008$, in comparison with our
  previous best estimate of $1.057$. We adopt noise variance
  multipliers of $1.017$ and $1.071$ for the polarization and total
  intensity observations, respectively, and an uncertainty of $0.9\%$.
\item {\bf Bright Sources:} The large ($\sim 143 \, {\rm deg^2}$)
area covered by CBI makes it impossible to completely avoid bright
radio sources: there are two $S_{30} > 300 \, {\rm mJy}$ sources
in the fields. For the observed distribution
of sources in flux density ($N(>S) \sim S^{-1}$), the brightest sources (or constant
fractional residuals to them) will dominate the map variance. To measure
the possible effect of these sources we reanalyzed the data after
removing all individual CBI pointings where a point source with an
apparent flux density $S_{30} > 80 \, {\rm mJy}$ was evident, which
removed 33 out of 259 pointings. This reduces the power spectrum by
less than $15 \, {\rm \mu K^2}$ averaged over all $\ell$ and shows no characteristic
trend of increasing to high-$\ell$, which would be expected of residual
source contamination.
\item {\bf Source Structure:} We project pure point source templates
  for the NVSS sources.  The NVSS resolves $\sim 20$ percent of the
  sources it detects.  If the sources are resolved by the CBI, then we
  would expect leakage into the power spectrum.  Using
  Montage\footnote{http://montage.ipac.caltech.edu} we mosaic the NVSS
  4 deg x 4 deg maps to cover each of our CBI fields, and, zeroing out
  all pixels below 3 times the NVSS RMS noise level, simulate CBI
  observations of those maps.  These observations are run through the
  full CBI data reduction pipeline, projecting out sources using the
  NVSS catalogs above 3.4 mJy.  We find that our projection method
  removes $>$99.9\% of the source power, with the residual power a
  factor of 20 times lower than the observed CBI high-$\ell$ signal.
  This is a conservative estimate of the signal from source
  extendedness, since the high-frequency emission from AGNs tends to
  preferentially come from compact cores.  In addition to testing the
  impact of source size, this is a powerful test of our data pipeline
  and the quality of the NVSS catalog.

\end{enumerate}

\section{Results \& Interpretation}
\label{sec:results}

\subsection{Power Spectrum}
\label{subsec:powerspectrum}

\begin{figure}
\centering
\includegraphics[width=16cm]{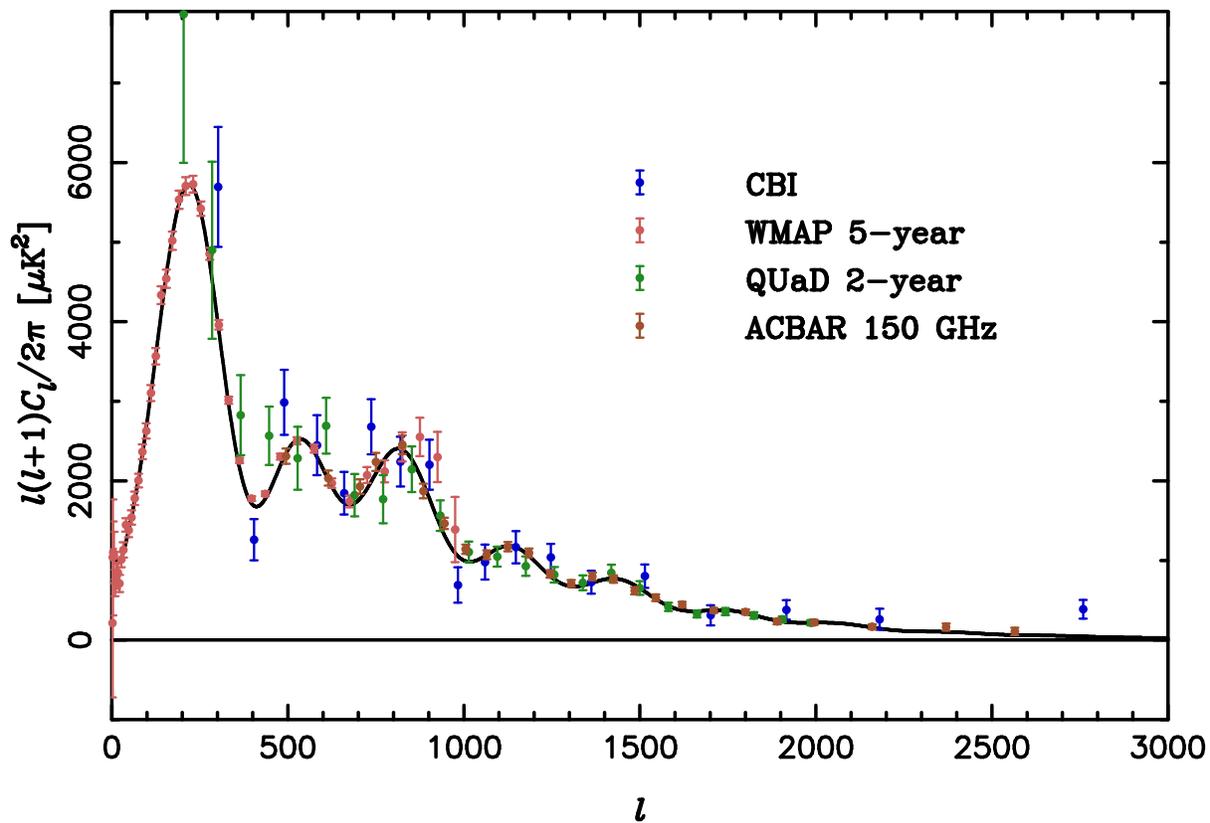}
\caption{CBI total intensity power spectrum.  The blue points are the
  CBI power spectrum in this work, given in text form in
  Table~\ref{tbl:powspec}.  The salmon points are the WMAP 5-year
  spectrum \citep{Nolta08}.  The green points are the QUaD 2-year
  spectrum \citep{quad08}.  The burnt sienna points are the ACBAR 150 GHz
  spectrum \citep{acbar08}.  The full CBI spectrum, including bin
  correlations and window functions, is available online. }
\label{fig:newspec}
\end{figure}

The final CBI total intensity power spectrum is shown in
Figure~\ref{fig:newspec}, together with the WMAP 5-year, ACBAR, and QUaD power
spectra, and best-fit tilted $\Lambda$CDM power spectrum model.  The spectrum
is given in Table~\ref{tbl:powspec}, and full window functions and
bin-bin correlations are available on-line.\footnote{http://www.astro.caltech.edu/$\sim$tjp/CBI/data/index.html}

\begin{figure}
\centering
\includegraphics[width=16cm]{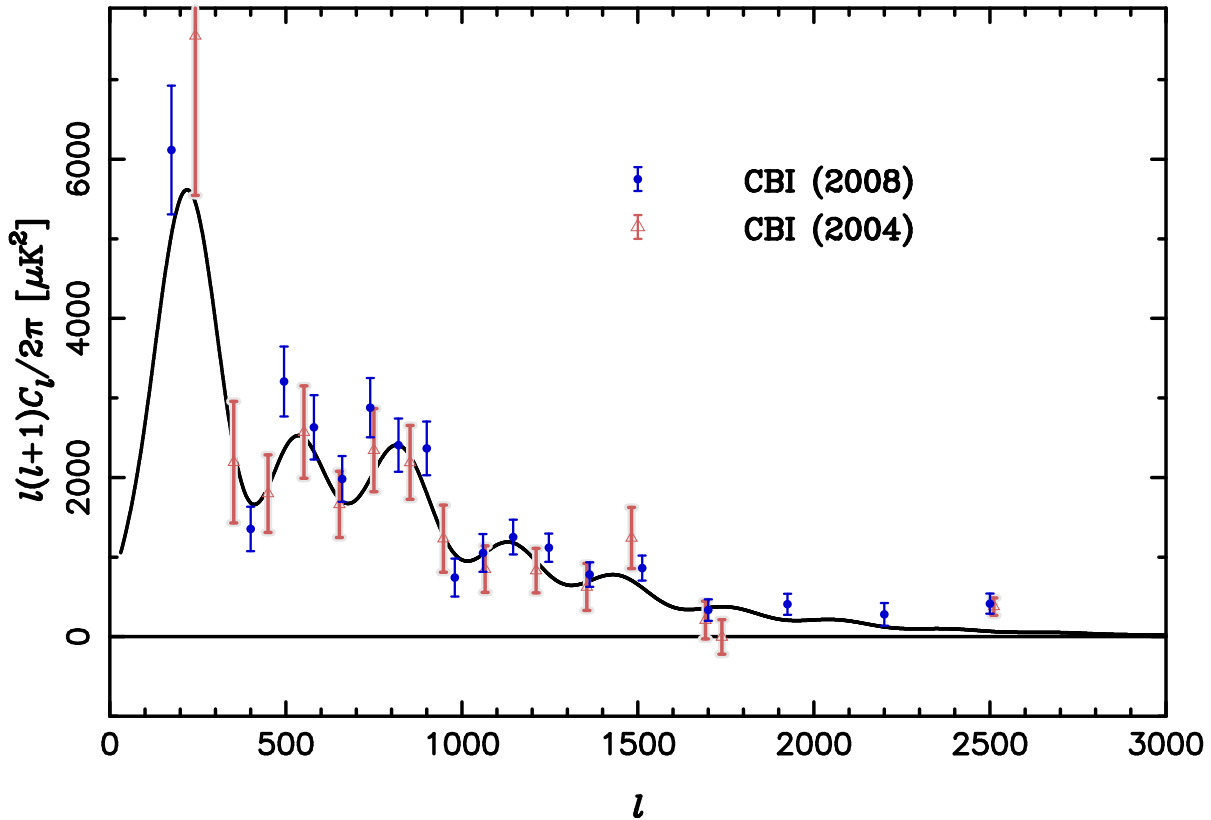}
\caption{A comparison of the CBI 5-year power spectrum, including
results from the GBT 30 GHz survey, with the two-year CBI power spectrum of
\citet{paper7}.  Note the changed finer binning at high $\ell$ for the
power spectrum from the 5-year data.  The window function of the
\citet{paper7} highest-$\ell$ bin extends from $\ell \sim 2000$ to
$\ell \sim 3500$.  Note that the error bar on that very big bin is
about the same as for the last of the finer bins, in spite of the
large \citet{paper7} bin being broken up into three distinct bins in
this work.  The main reasons for the improvement in the spectrum is
the factor of two more data, and the development of analysis
techniques that allowed us to combine these disparate datasets.  In
the damping tail, the new spectrum is about a factor of two
improvement over \citet{paper7}.
}

\label{fig:oldnew}
\end{figure}

A comparison of these results with the previous CBI power spectrum is
shown in Figure~\ref{fig:oldnew}.  The marked improvement is due to
several factors: the inclusion of $\sim 50\%$ more data, use of the
GBT 30 GHz observations to reduce the uncertainty due to point
sources, and the algorithmic
improvements described in \S~\ref{subsec:pipeline}.

\begin{figure}
\centering
\includegraphics[width=16cm]{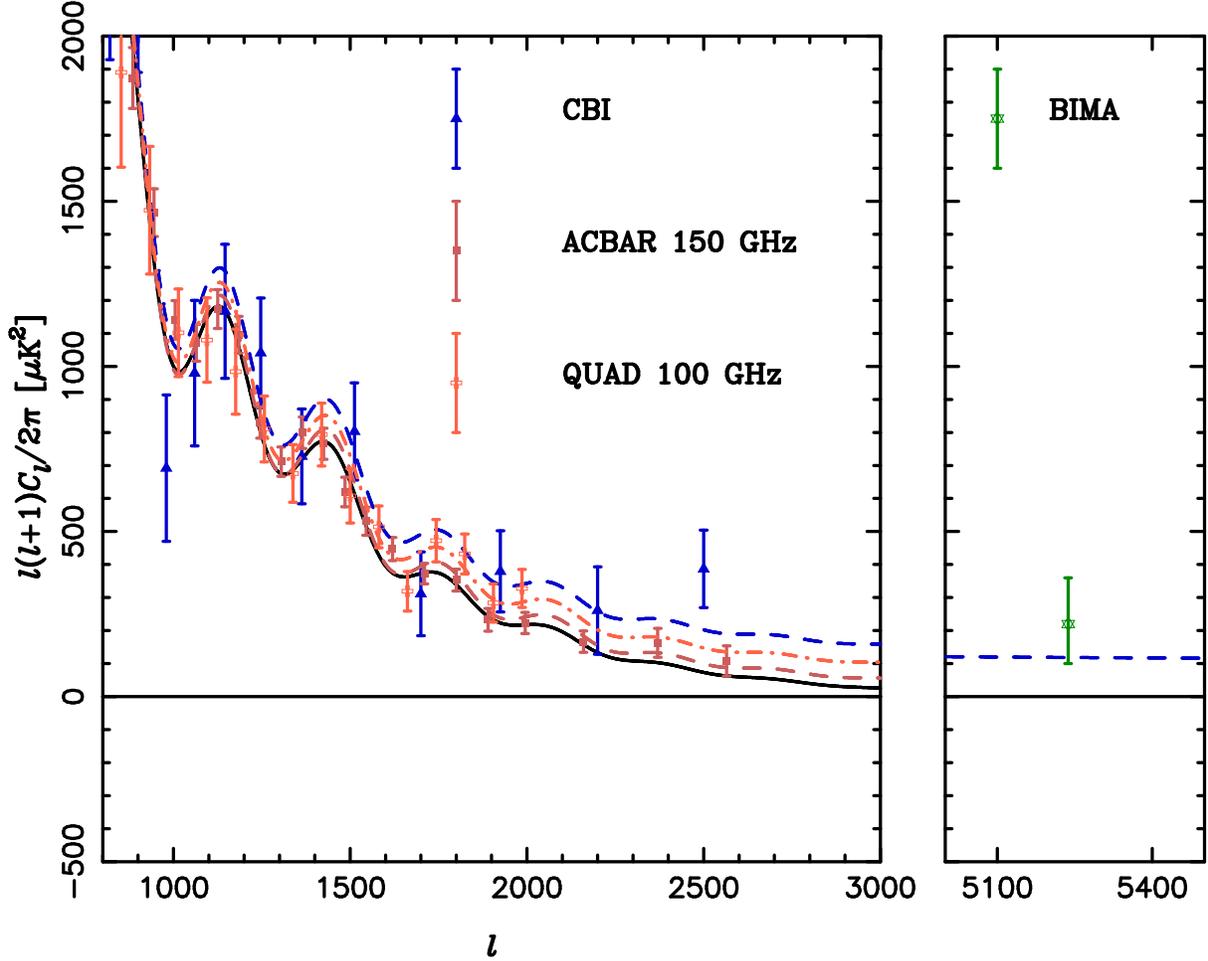}
\caption{The CBI, ACBAR, QUaD, and BIMA power spectra at small angular
  scales are contrasted.  The solid black line shows the
  tilted $\Lambda$CDM model from Section~\ref{subsec:powerspectrum} for the CMB primary anisotropies. 
  The dashed lines include the contribution of secondary SZ
  anisotropy using the model of \citet{komatsuandseljak} with the
  best-fit template scaling of $3.5$ (in bandpower) that we have
  determined using only the CBI data.  The SZ plus primary anisotropy
  power combination at 30 GHz is the blue-dashed line.  Note that,
  apart from fitting the CBI power spectrum, it passes through the
  BIMA point at $\ell \sim 5300$ \citep{dawson06}.  We have also
  forecast the level for SZ plus primary anisotropies at 150 GHz
  (red dashed line)  and  100 GHz (orange dot-dash line).  These are
  compared with the power spectra of ACBAR \citep{acbar08}and QUaD \citep{quad08}. } 
\label{fig:highell}
\end{figure}

At $\ell > 1800$ there remains a clear excess of power over the
expected intrinsic CMB anisotropy. This is shown in
Figure~\ref{fig:highell} along with ACBAR \citep{acbar08}, QUaD
\citep{quad08}, and BIMA
\citep{dawson06} measurements. To quantify this we use the
\citet{komatsuandseljak} analytic and the \citet{paper6} SPH
simulation-based predictions of the SZ 
angular power spectrum.  (See Section~\ref{subsec:parameters} for more
details.)   An amplitude for the SZ spectrum is then fit in
conjunction with an amplitude for the intrinsic anisotropy spectrum,
using a canonical tilted-$\Lambda$CDM model\footnote{Flat,
  $\Omega_bh^2=0.0223$, $\Omega_{CDM}=0.108$, $\tau=0.087$, $n_s=0.96$,
  and including the effects of gravitational lensing.  No SZ
  contribution is included, as that is handled separately.} as a shape.  The
templates are 
calculated with $\sigma_8=0.8$ and $\Omega_bh=0.0321$.  This results in a
$3.1\sigma$ detection of power in excess of the intrinsic anisotropy (a
best-fit scaling of $3.5 \pm 1.3$ of the nominal KS template).  The SPH
template has a best-fit scaling of  $5.4 \pm 1.8$  and a $3.4\sigma$
detection significance. 
The detection significances are calculated using $\sqrt{-2\delta \log
  \left ( \mathcal{L} \right )}$, where the likelihood difference is
between the best-fit CMB-only shaped model and the best-fit
CMB+template model.  We  
show the total power spectrum including the contribution of the
best-fit SZ spectrum by the {\it dashed} lines in
Figure~\ref{fig:highell}.  Of note is the marked change in the
magnitude of the SZ power spectrum between 30 and 150 GHz, and the
fairly broad contribution of clusters down to as low as $\ell \sim
1000$.

\begin{figure}
\centering
\includegraphics[width=16cm]{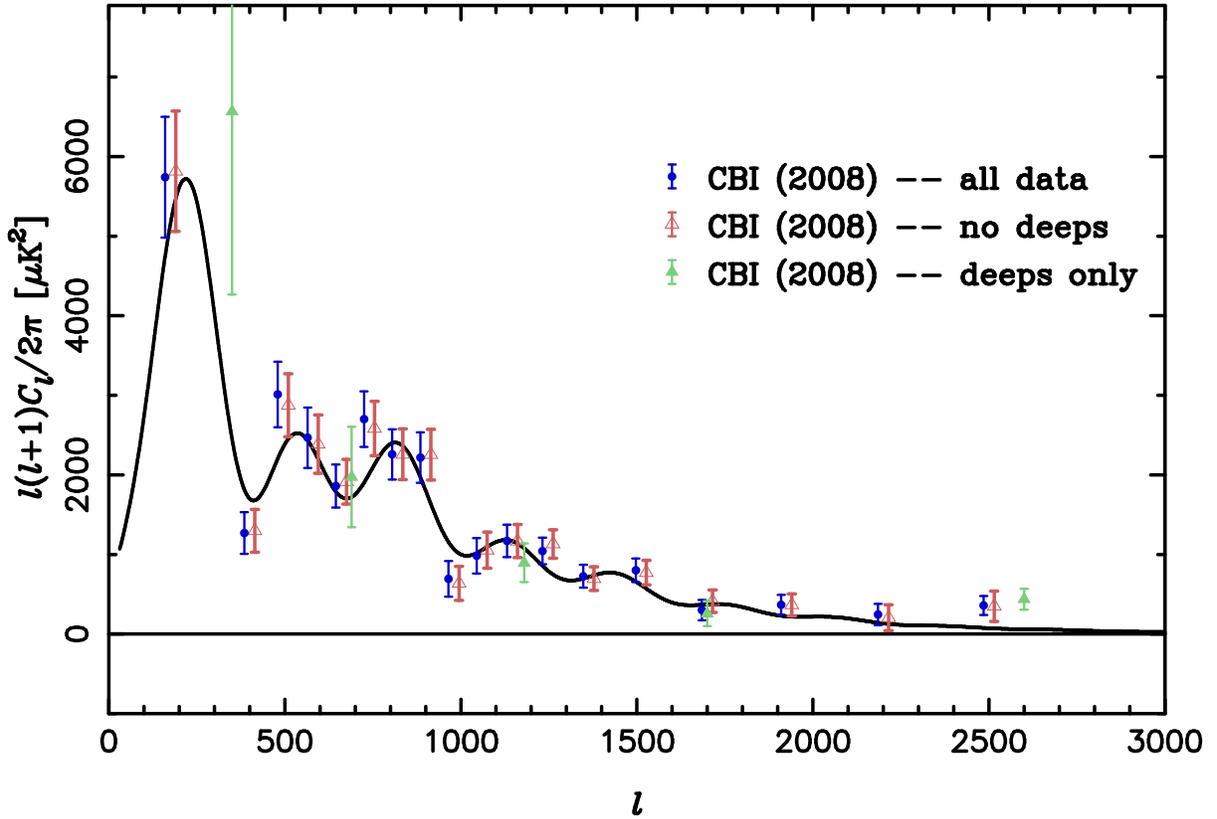}
\caption{A comparison of the power spectrum obtained from all CBI TT data with that obtained
when the deep fields are excluded, and with that obtained when only
the deep fields are used.  Although the error for the no-deeps at
$\ell \sim 2500$ is larger, and consistent with no excess at the $\sim
1-\sigma$ level, the overall mean amplitude is about the same as for
the deep-only case.  The all-data and no-deeps spectra are at
the same $\ell$, but have been offset for clarity.
}
\label{fig:allvsnodeep}
\end{figure}

Since the secondary SZ anisotropy will be highly non-Gaussian,
uncommon structures in the $\sim 4 \, {\rm deg^2}$ of ``deep'' CBI
pointings could bias the power spectrum (although the statistical
weight of the wide, shallow area is about $2.5$ times of that of the
deep data).  We re-ran the power spectrum extraction both {\it excluding}
the deep field data from the analysis, and using {\it only} the deep
data.  The results are shown in  
Figure~\ref{fig:allvsnodeep}. While the error bars are increased by
$\sim 25\%$, the amplitude of the power spectrum at $\ell > 1000$ is
not reduced, indicating that the small-scale excess power is not a
peculiar property of the deep fields observed by CBI.  An assessment
of the cosmic variance in the high-$\ell$ CBI power spectrum, under
the assumption that the dominant signal over intrinsic anisotropy is
due to SZ clusters, is presented in \S~\ref{subsec:parameters}.

\begin{figure}
\centering
\includegraphics[width=16cm]{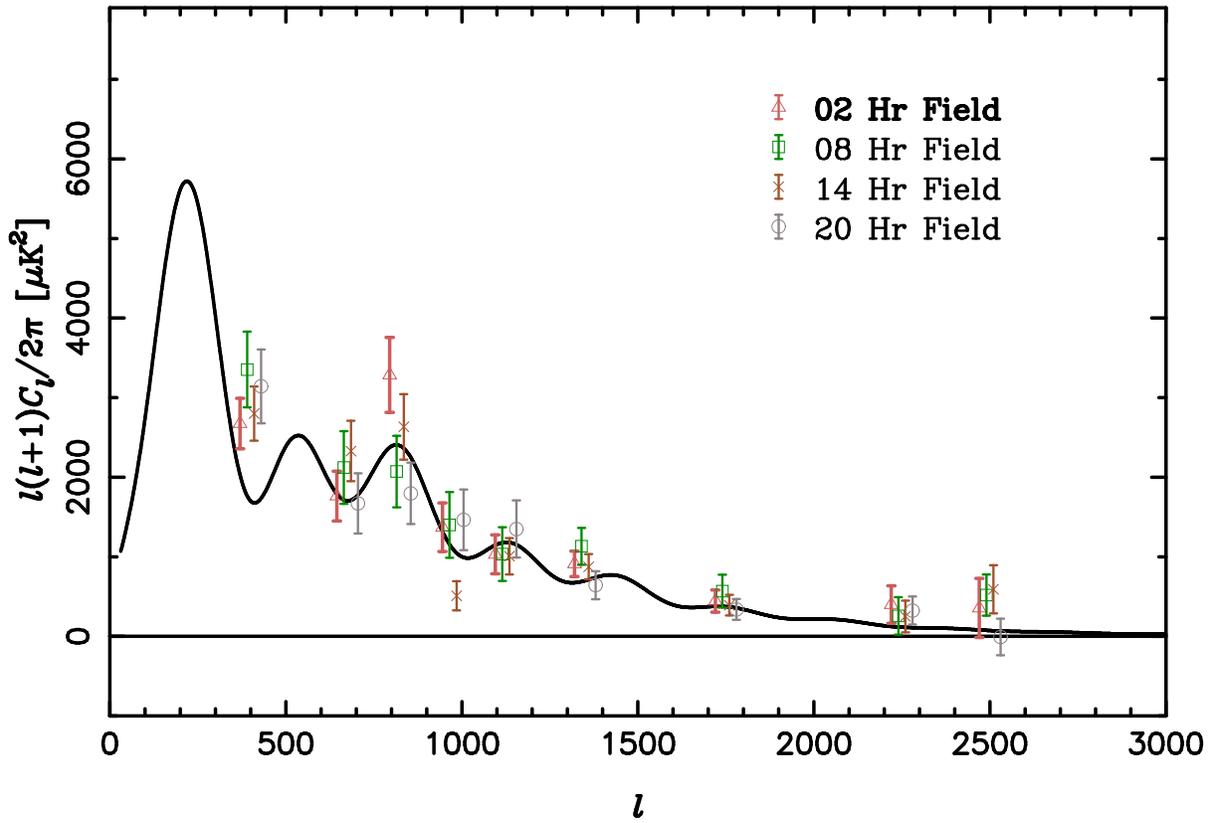}

\caption{Power spectra from individual CBI fields show how the
  bandpowers fluctuate from field to field.  The individual fields are
  all consistent with each other, and each sees power above the CMB.
  The spectra are staggered in $\ell$ for clarity in plotting.}
\label{fig:individualps}
\end{figure}

Power spectra from the individual CBI fields are shown in
Figure~\ref{fig:individualps}. To increase the signal-to-noise ratio a
coarser binning has been used.   Each of the CBI mosaics shows an
excess of at least 0.7$\sigma$ , with levels of ( $2.7 \pm 2.6$,
$7.6\pm 2.9$, $2.6\pm 2.3 $, and $1.6 \pm 2.2$ ) above the CMB using
the $\sigma_8=0.8$ KS template for the (02, 08, 14, 20)-hour 
mosaics.  The individual fields are all consistent with the same
excess level (reduced $\chi^2=1.01$ per dof), with the most discrepant
field 1.6$\sigma$ away from the mean.  

The CBI data by themselves cannot determine the source of the excess.
In particular, they cannot distinguish an SZ foreground from point
sources (if one fits for a point source and an SZ amplitude
simultaneously, the power spectra are sufficiently similar that the
error bars more than triple).  If instead of an SZ template, we model the excess with a
point source-like template, we find its total amplitude is $0.180 \pm
0.052 \jysq$ above the expected contribution from unresolved sources
of $0.046 \jysq$.   To high accuracy, the best-fit SZ excess level is
linearly dependent on the input source level.  For the KS template,
the best-fit SZ amplitude is $q_{sz}$=4.59-24.7$\times q_{iso}$ where $q_{sz}$ is in units of the predicted
SZ signal for $\sigma_8=0.8$ and $q_{iso}$ is the faint source
contribution in $\jysq$.  One can use this relation to ask how
different mean source levels would affect the excess:  a value of
0.186 for $q_iso$ would make the high-$\ell$ excess disappear, a value of 0.146
would make the best-fit CBI value be equal to the predicted KS level,
and a value of 0.096 would make the best-fit CBI value be within
1-$\sigma$ of the predicted KS level. 


\subsection{Diffuse Foregrounds}
\label{subsec:foregrounds}

While on small scales, point sources are the largest 30 GHz
foreground, on larger scales ($\gta$ 1 degree), diffuse foregrounds dominate.
At frequencies of $\sim 30$~GHz, the major known diffuse Galactic foregrounds are
synchrotron radiation and free-free emission.  Vibrational (thermal) dust
emission is negligible at frequencies below $\sim 50$~GHz. However, there
is considerable evidence for an additional component which is closely
correlated with FIR dust maps and that appears to dominate the spectrum in
the range $\sim 10-50$~GHz
\citep{Leitch97,Banday03,deOliveira-Costa04,Finkbeiner04a,Davies06,Bonaldi07,Hildebrandt07,Dobler08a,Dobler08b}.
Both the COBE-DMR and WMAP datasets have shown that the high latitude
sky is dominated by such a foreground closely linked with dust emission,
usually traced in the FIR ($\lambda \sim 100~\mu$m). The most popular
candidate for this anomalous component is the so-called ``spinning
dust'' emission \citep{Draine98a, Draine98b,Haimoud08}, but the situation
is far from clear.

Whether or not the dominant diffuse emission is spinning dust or
flat-spectrum synchrotron is not of great importance in the context of
this paper. However, we need to quantify the level at which diffuse
foregrounds are present in our data. We use the fact that the FIR maps
are a good tracer of the foreground morphology at these
frequencies. In the WMAP data at K and Ka-bands (23 and 33~GHz,
respectively), there is a strong correlation with the 100~$\mu$m
\citet{Schlegel98} map. If the emissivity of the dust emission is roughly
constant for a given region, cross-correlation between the FIR map and
CBI data provides a very sensitive method to detect such
emission. Furthermore, IRAS data have adequate resolution to cover the
angular scales measured by CBI.

We simulated CBI data based on the \citet{Schlegel98} $100~\mu$m map as our
foreground ``reference'' map. The $100~\mu$m maps were converted to CBI
visibilities using the {\sc mockcbi} software based on the real
observed visibility data sets. To convert to the approximate signal
levels expected at 30~GHz, we took a typical dust emissivity at 
high Galactic latitude of $10~\mu$K/(MJy/sr)
\citep{Banday03,Davies06}. Although the emissivity can vary by a
factor of a few over the sky, it provides an initial guess for the
amplitude of the signal \ie\ we expect a cross-correlation coefficient
of $\sim 1$ based on previous data on larger angular scales. We also
repeated the analysis using H$\alpha$ data from the compilation of
\citet{Finkbeiner03}. H$\alpha$ is known to be good tracer of free-free
emission at high Galactic latitudes where dust extinction is small
\citep{Dickinson03}.  We scaled the H$\alpha$ template by
 5.83~$\mu$K per Rayleigh, which is the expected value at 31GHz,
assuming $T_e\approx 8000$~K \citep{Dickinson03}.

The simulated foreground visibilities were fitted to the CBI data
using the template-fitting method of Appendix \ref{sec:templates}. For the  
dust template of \citet{Schlegel98}, we found a combined correlation
coefficient of $1.18 \pm 0.53$ (2.2$\sigma$) for the 
entire dataset. This suggests that a very small level of contamination
from Galactic emission might exist in the CBI data, at a level similar
to those observed at larger angular scales
\citep{Banday03,Davies06}. The fluctuation power $C_{\ell}$ in the FIR dust
scale as $C_{\ell} \propto \ell^{-3}$, \citep{Gautier92}, thus the power
will be mostly at large angular scales. For the H$\alpha$ template we
find  a combined correlation coefficient of $0.78 \pm 0.58$.  For both
foreground templates, we find a negligible impact on the power
spectrum, with the high-$\ell$ excess changing by less than  1\%.
The emissivity factor is close to what is expected at high 
latitudes and indicates that the CBI fields are relatively low in
foregrounds on angular scales $<1^{\circ}$. 
This is also supported by the field-by-field data splits (\S
\ref{subsec:powerspectrum}):  all the independent fields give
consistent power spectra.

\subsection{Cosmological Parameters}
\label{subsec:parameters}

We can place constraints on the standard parameters of tilted $\Lambda$CDM
cosmology by fitting a range of model spectra to the observed CMB (and
other) data.  Model spectra were generated by CAMB
\citep{Lewis:1999bs} and the parameter constraints determined by a
modified version of the Monte-Carlo Markov Chain code, COSMOMC
\citep{cosmomc}.  For all parameter runs, we take into account the
effects of CMB lensing, and assume a prior that the universe is
spatially flat.  To capture
the non-Gaussian nature of the CBI's power spectrum, we use the
offset-lognormal approximation of \citet{Bond98} in parameter
analysis.  To convert model spectra to predicted CBI bandpowers, we
evaluate the window functions with a spacing of $\delta \ell=20$, and
use cubic Hermitian polynomial interpolation between the measured points.

We extend the treatment of the SZ foreground relative to previous
analyses \citep{paper6,paper7}.  We assume the angular power spectrum
from clusters scales in a simple analytic fashion:  $\propto
\sigma_8^7 \left (\Omega_b h \right )^2$ \citep{paper6,komatsuandseljak}.  We have
explored two sets of SZ templates: 
\begin{enumerate}
\item Semi-analytic templates from \citet{komatsuandseljak} (the \KS\ template), updated
to include the effect of $\sigma_8$ on the shape of the SZ power spectrum (E. Komatsu,
private communication).  The dependence of the shape on $\sigma_8$ is found
to be minimal.
\item Power spectra from tree-SPH simulations described in
  \citet{paper6} (the \SPH\ template).
\end{enumerate}
The \SPH\ template has relatively less power than \KS\ for fixed
cosmological parameters, and rises more quickly 
with $\ell$.  Unless stated otherwise the \KS\ template is used in all
parameter analysis.  Also unless otherwise 
indicated, all runs use the full $\ell$ range of data and explicitly
model the SZ foreground.  Based on a comparison of these theoretical
templates, we estimate that for a given background cosmology there is
a factor of $\sim 2$ uncertainty in the SZ power spectrum. This
corresponds to a $\sim 10\%$ systematic uncertainty in $\sigma_8$.
Because of this we do {\it not} link the SZ spectra to the background
cosmology, {\it e.g.}, through $\Omega_b$ or $\sigma_8$. Rather we use
an independent parameter $\sigma_8^{SZ}$ to describe the amplitude of
the SZ power spectrum and to index the family of shaped SZ power
spectra.

Table~\ref{tbl:wmap5cbiparams} shows the marginalized individual
parameter results for an analysis including WMAP 5-year
\citep{Nolta08, Dunkley08} and CBI data, using both SZ templates. This
includes a marginalization over the uncertainty in the residual point
source power spectrum using the distribution determined from the
simulations in \S~\ref{sec:pointsources}. 
Table~\ref{tbl:wmap5cmballparams} contains the same parameters, but with
the addition of more CMB data to CBI and WMAP: ACBAR \citep{acbar08},
BIMA \citep{dawson06}, VSA \citep{Dickinson:2004yr}, Boomerang
\citep{Montroy:2005, Jones:2005}, QUAD \citep[ adopting their ``pipeline
  1'' spectrum]{quad07}, and both CBI and DASI polarization data \citep{paper9,
  Leitch04}.  We henceforth refer to this data combination as CMBall.
The choice of SZ template has an impact on the value of $\sigma_8$
inferred from the primary fluctuations - the \citet{paper6} template
gives $\sigma_8$ values that are systematically higher by about 0.015
than the \citet{komatsuandseljak} template.  This is because the
\KS\ template is flatter than the \SPH\
template, so for fixed observed power at $\ell \sim 2000$,
\KS\ will remove \textit{more} power in the region
of the third peak, dropping $\sigma_8$.  Our values of $\sigma_8$ are
also lower than those in \citet{Dunkley08} where the level of the
\KS\ template is capped at twice the level
predicted for a $\sigma_8=0.8$ universe.  We allow the amplitude to
float freely, and the CBI data set the level to be around 2.5 times
the $\sigma_8=0.8$ prediction, again resulting in more power being
removed from the third peak and a lower primary $\sigma_8$ value.

\begin{figure}
\centering
\includegraphics[width=8cm]{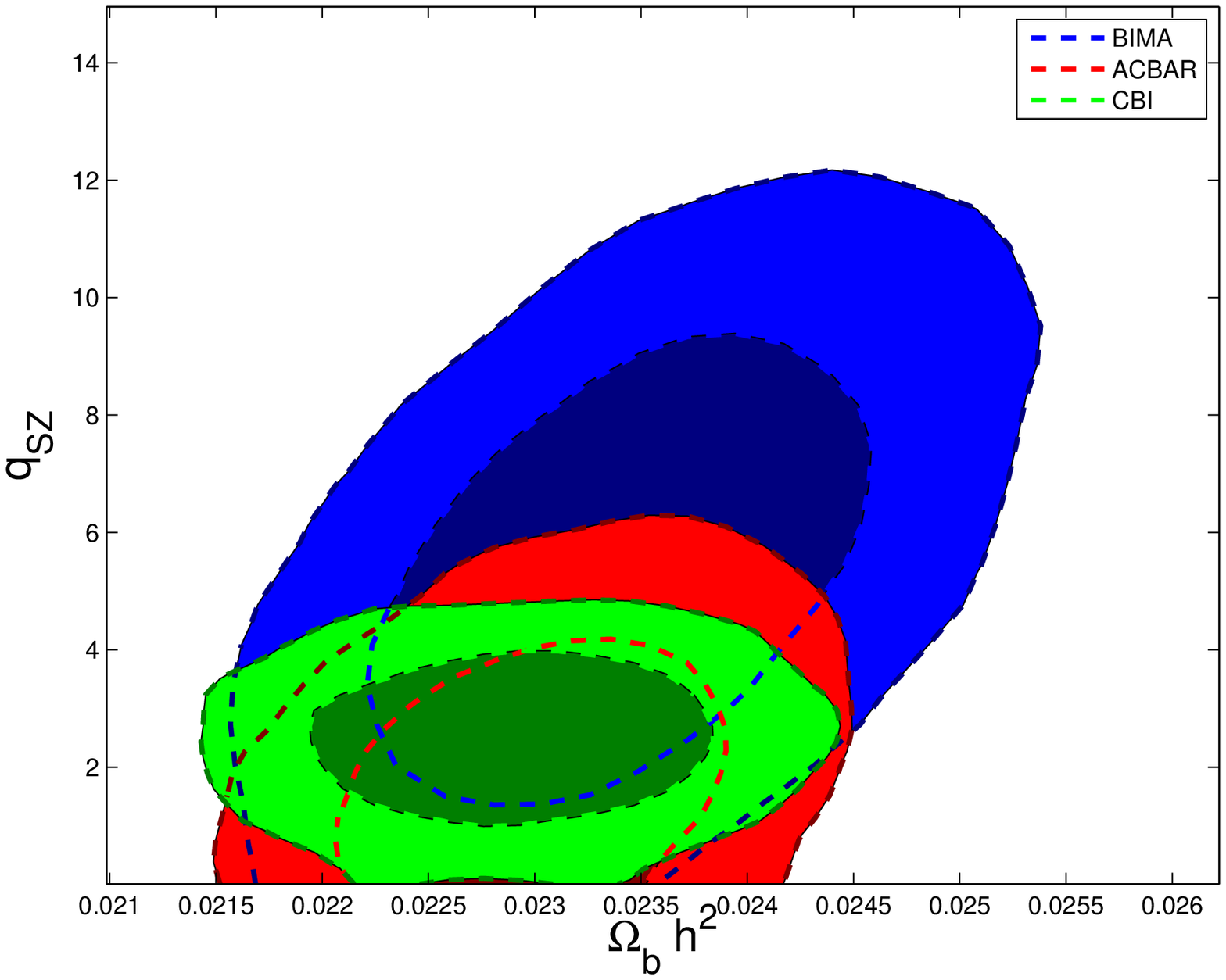}
\includegraphics[width=8cm]{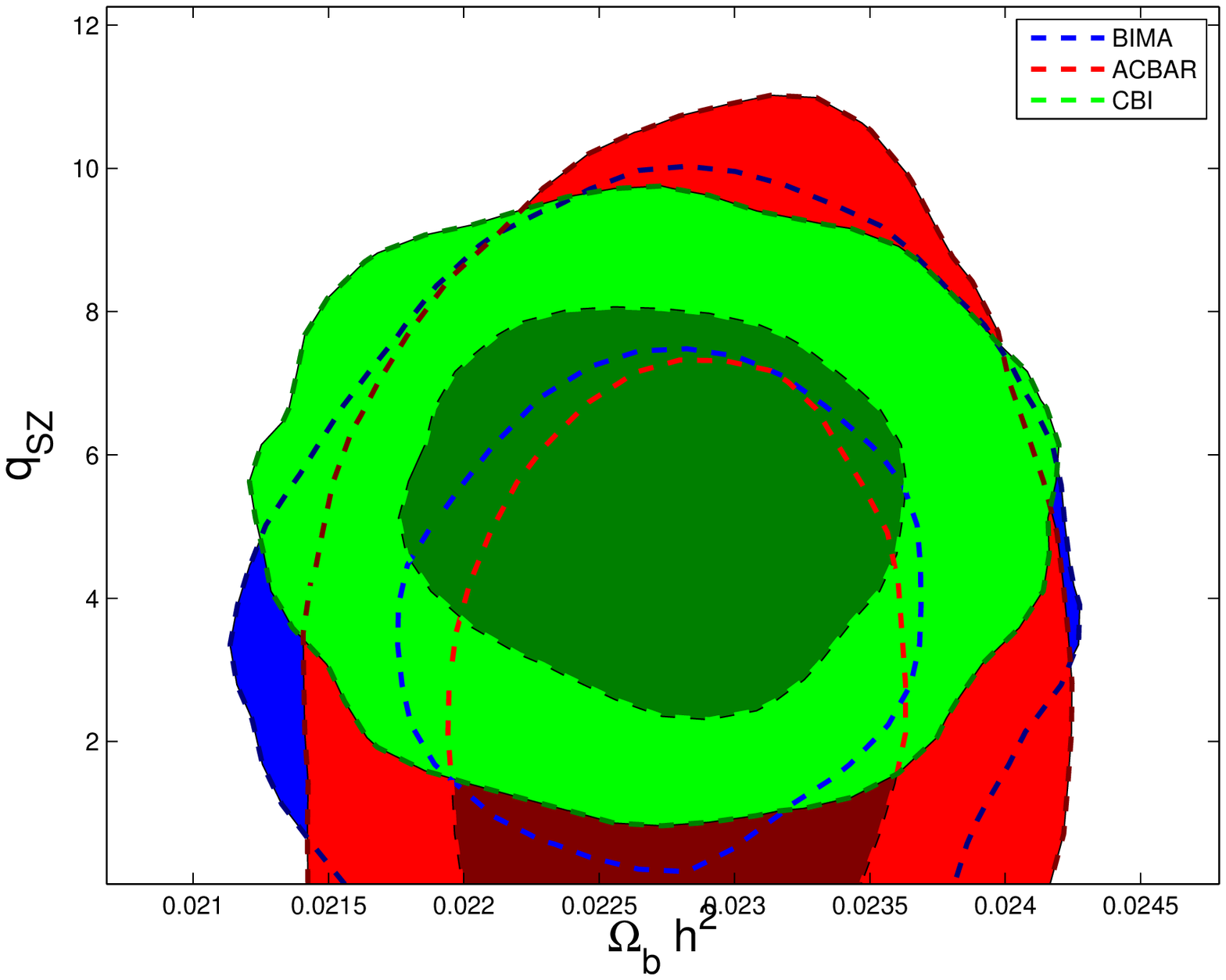}
\caption{1- and 2-$\sigma$ likelihood contours of the SZ amplitude and the baryon
  density, as determined from MCMC chains for CBI, ACBAR, and BIMA, as
  indicated.  We ran chains for WMAP5 plus, in turn, CBI, ACBAR, and BIMA, fitting an SZ excess
  template to each case.  The left panel shows the results when using the \KS\ template,
  and the right panel shows the same using the SPH SZ template,
  marginalizing over the other parameters.  For both
  templates, the 1-$\sigma$ regions of the three
  experiments are all in excellent agreement if the excess is due to
  SZ.}
\label{fig:3expsz}
\end{figure}

The values of $\sigma_8$ inferred from the high-$\ell$ power spectrum
are $\sigma_8^{SZ}=0.910 \pm 0.064$ for CBI+WMAP5 and
$\sigma_8^{SZ}=0.922 \pm 0.047$ for CMBall. As a point of comparison,
using the \SPH\ template, we find $\sigma_8^{SZ}=1.015 \pm 0.060$ for
CBI+WMAP5 and $\sigma_8^{SZ}=0.977 \pm 0.049$ for CMBall.  CMBall has
a (slightly) higher inferred $\sigma_8^{SZ}$ than CBI+WMAP5 for the
\KS\ template, whereas it has a lower one for the \SPH\ template. This
is due to the BIMA data falling above the CBI prediction for the
relatively flat \KS\ template, while they fall below the CBI
prediction for the \SPH\ template.  We find that CBI, ACBAR, and BIMA
are consistent with each other if the high-$\ell$ excess is due to SZ
- see Figure~\ref{fig:3expsz} for full MCMC chains comparing the three
experiments using both SZ templates.  For comparison with other
experiments, we give the best-fit CBI excess level (from
Table~\ref{tbl:wmap5cbiparams}) for the two templates at selected
$\ell$ and frequencies.  The results are summarized in
Table~\ref{tbl:excesspred}.

When we extend the $\Lambda$CDM
model to include tensor modes, we find no detection of them:  $r<0.32$
(95\%) for CMBall, where r is the tensor-to-scalar power ratio at the pivot
wavenumber 0.05 $h$ Mpc$^{-1}$.  

\begin{figure}
\centering
\includegraphics[width=16cm]{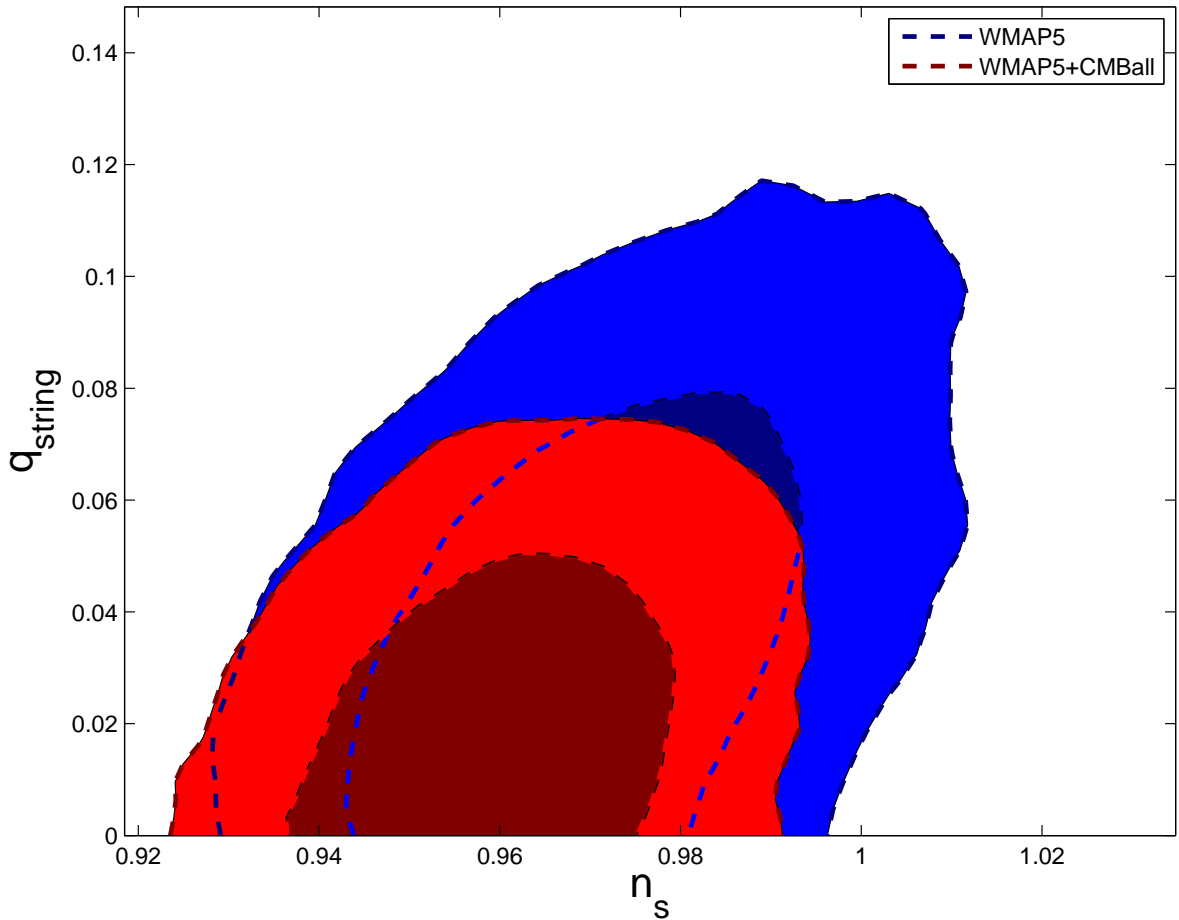}
\caption{1- and 2-$\sigma$ likelihood contours of cosmic string amplitude and
  $n_s$, marginalized over other parameters.  The string amplitude is
  relative to the $\mathcal{C}_{\ell}$-template of \citet{Levon08},
  which is normalized to a string
  tension of $G{\mu}=1.1\times 10^{-6}$.  With only WMAP5 data, $n_s$ is
  partially degenerate with $q_{string}$, and $n_s < 1$ is no longer
  significant at a 2-$\sigma$ level.  The addition of the high-$\ell$ CMB data 
  breaks this degeneracy, and $n_s<1$ at the 2-$\sigma$ level holds
  even with the addition of cosmic strings to the parameter analysis.}
\label{fig:qstring}
\end{figure}

We have explored the potential impact of cosmic strings in the CMB
using a string template of \citet{Levon08}.  We treat the string
contribution as a template of fixed shape added to the power spectrum, and allow its
overall amplitude $q_{string}$ to float, just as we have for the SZ
templates with overall parameter $q_{SZ}$.  We re-emphasize the 
\citet{Levon08} caution that the power spectrum from 
strings is not uniquely determined and its shape depends on the details
of the string properties.  While no combination of CMB data detects cosmic
strings, the addition of high-$\ell$ CMB data adds a significant
constraint to the maximum allowed string amplitude: the 95\% string
upper limit from CMBall is 65\% of the limit from WMAP5 alone (see
Fig. \ref{fig:qstring}).  The 
high-$\ell$ data constrain strings because the power spectrum from
them generically falls off more slowly with $\ell$ than that from the
adiabatic fluctuations since the string fluctuations are not subject
to Silk damping \citep{Levon08}.  The amplitude $q_{string}\propto
\left ( G\mu \right )^2$ can be expressed in terms of the string
tension $\mu$ times Newton's constant $G$.  The limits on $G\mu$ (which
scales like the square root of the power spectrum amplitude) are
$G\mu<3.4\times 10^{-7}$ for WMAP5 only, and $G\mu<2.8\times
10^{-7}$ for CMBall.  With WMAP only, the string amplitude is highly
degenerate with $n_s$ (\eg\ \citet{Battye06}, see Fig. \ref{fig:qstring}), and there is no
longer a ``detection'' of $n_s<1$, rather we obtain $n_s= 0.972 \pm 0.018$ with a 95\%
upper limit of 1.007.  The addition of the high-$\ell$ CMB breaks this
degeneracy, and $n_s$ remains less than one with high significance:
we obtain $n_s = 0.961 \pm 0.015$ with a 95\% upper limit of 0.990.

\begin{figure}
\centering
\includegraphics[width=16cm]{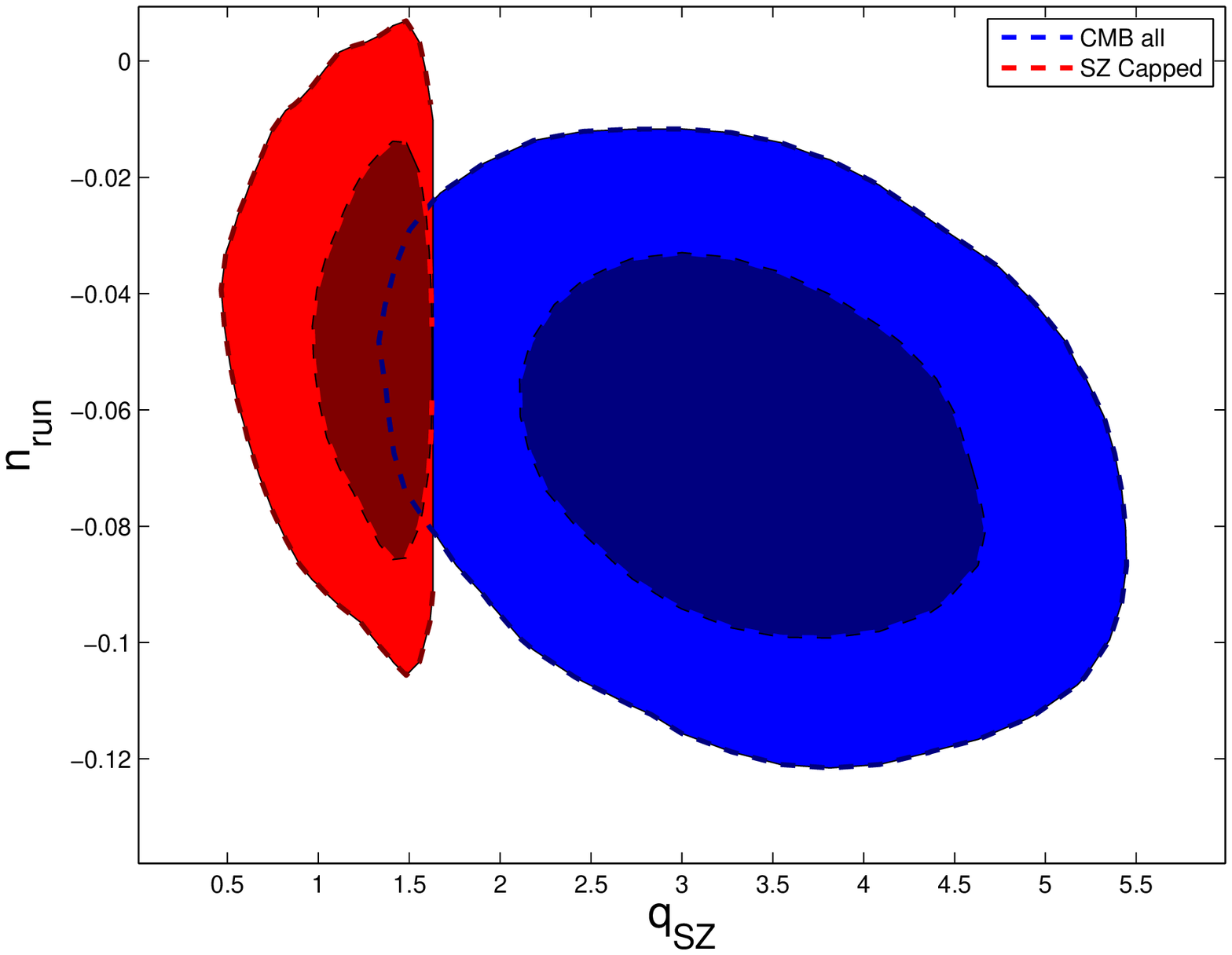}
\caption{ 1- and 2-$\sigma$ likelihood contours of the running of the spectral
  index $n_{run}=\nrun ( 0.05 h$ Mpc$^{-1})$, and  SZ amplitude
  $q_{SZ}$, marginalized over other parameters.  A high level
  of the SZ template pulls power out of the primary CMB fluctuations
  in the region of the third peak.  This drives $\nrun$ more negative.
  The artificial cropping of $q_{SZ}$ at a value of 2.0 as used in the
  WMAP5 analysis of \citet{Dunkley08} clearly distorts the picture
  relative to a freely floating $q_{SZ}$.}

\label{fig:nrun}
\end{figure}

We find that the addition of the CBI data have a strong impact on the running
of the scalar spectral index, $\nrun$ (evaluated at a pivot wavenumber
of 0.05 $h$ Mpc$^{-1}$).  For the KSSZ template, we find $\nrun=\left (
-0.041\pm 0.031, -0.048\pm 0.028, -0.066 \pm 0.022 \right )$ 
for ( WMAP5, WMAP5+CBI, CMBall), respectively.  For the SPH SZ
template, we find $\nrun = ( -0.039 \pm 0.030, -0.042 \pm 0.027,
-0.059 \pm 0.022)$ for ( WMAP5, WMAP5+CBI, CMBall), respectively.
Thus, with CMBall, the running of the spectral index is a 2.76-$\sigma$ ``detection''
for the SPH template, and 3.03-$\sigma$ one for the KS template.  The
detection is substantially driven by the high values obtained for the SZ
signal.  When the allowed SZ level is capped at twice the nominal KS
value, we find that limits on $\nrun$ for CMBall are $ -0.048\pm
0.021$  and $ -0.043 \pm 0.021$ for the KS and SPH templates, around
2-$\sigma$ detections (see Fig. \ref{fig:nrun}).  We stress
that the inferred value of $\nrun$ depends not only on the amplitude
of the SZ signal, but somewhat on its shape as well, hence the difference in
$\nrun$ between the two templates.

Since clusters are compact sources, the distribution from field to
field will not be Gaussian.  We have assessed the effect on the sample
variance in the SZ cluster power spectrum for the CBI coverage region
using the simulated maps of \citet{whiteplanck}.  These consist of 10
maps of the thermal SZ effect, each $10^{\circ}\times 10^{\circ}$ in
size.  These are calculated using large N-body simulations in a
$\Lambda{\rm CDM}$ cosmology with $\sigma_8=1$.  The pressure profiles
of the clusters were made assuming the gas density follows the dark
matter density, and the temperature is isothermal and proportional to
$M_{halo}^{2/3}$ \citep{whiteszsim}.  The SZ maps are line-of-sight
integrations of these 3D pressure configurations.  Fake observations
of these fields are constructed using the real CBI $uv$ coverage and
per-visibility noise levels and they are run through our power
spectrum extraction. The thermal SZ power spectrum at $\ell > 1800$
emerging from this analysis has a mean level of $\sim 200 \, {\rm \mu
K^2}$. When only the CBI deep fields are used we find an RMS in the SZ
power spectrum of $\sim 100 \, {\rm \mu K^2}$, that is a fractional
scatter of about 50\% in the (noiseless) spectra.  When the full CBI
sky coverage is probed instead of just the deep fields, we find a
fractional scatter in the SZ spectrum to be 21\%.  The 21\% scatter
could be driven by non-Gaussian errors in the deep fields.  To test
this, we excluded the deep fields from the CBI mosaics and repeated
the analysis.  We found that the fractional scatter dropped to 19\%,
hence the deep fields do not drive up the variance of the total result
by a substantial amount.  We expect that the non-Gaussian component
will be quite sensitive to $\sigma_8$ since the number of clusters
drops dramatically as $\sigma_8$ is dropped, increasing the Poisson
fluctuations.  Thus the sample variance in the SZ spectrum depends
heavily on the value of $\sigma_8$, and since the sample variance
errors are highly correlated between multipole bins, these results
have not been fed into further analysis, but should be taken as
indicative.

\section{Summary}
\label{sec:summary}

We have presented the total intensity power spectrum resulting from
five years of dedicated CBI observations and campaigns of point source
foreground observations with the OVRO 40-m and the GBT.  The OVRO and
GBT data allow us to greatly improve our estimate of the power from
faint 30 GHz radio sources.  On its own, the CBI cannot distinguish
between power spectra from point sources and from other sources such as the
SZ effect from galaxy clusters, and so these supporting observations
are essential.  Our data support the existence of excess
power above the primary CMB on small angular scales at $\sim 3
\sigma$, and are the most sensitive constraints to date on the
statistical SZ cluster foreground.  In extensive testing, we find no
evidence of any instrumental systematic effect capable of giving rise
to the excess.  By running the NVSS maps (with all pixels below three
times the NVSS RMS noise zeroed out) through the CBI pipeline, we find
that our source treatment rejects $>$99.9\% of the power in known
sources.  This test confirms the quality of the NVSS catalogs, the
validity of the projection of known sources, and the insensitivity of
the CBI to extended sources.  If the excess is due to the SZ effect, we find that
other data, notably ACBAR and BIMA, are consistent with the excess
seen by CBI.  For two different SZ spectral templates, we find that
the inferred $\sigma_8$ from the excess is marginally inconsistent
with those derived from the primary fluctuations, $0.922 \pm 0.047$
and $0.988 \pm 0.049$ vs. $0.769 \pm 0.031$ and $0.784 \pm 0.030$.  To
determine definitively the implications of the observed small-scale
excess power further theoretical work is needed.  We find that
high-$\ell$ data break the degeneracy between the tilt of the spectral
index $n_s$ and potential contributions from cosmic strings.  From
running n-body simulations through the CBI pipeline, we find that the
scatter in the expected level of the signal from clusters due to their
non-Gaussian nature is about 20\%.

The CBI has been supported by funds from the National Science
Foundation under grands AST 9413935, 9802989, 0098734, and 0206416, by
the California Institute of Technology, by Maxine and Ronald Linde,
Cecil and Sally Drinkward, Barbara and Stanley Rawn Jr., Rochus Vogt,
the Canadian Institute for Advanced Research, and by the Kavli
Institute for Cosmological Physics.  We thank E. Komatsu for providing
us with additional tabulations of SZ power spectrum templates.  All
computations were performed on the Canada Foundation for Innovation
funded CITA Sunnyvale cluster.   Part of
the research described in this work was carried out
at the Jet Propulsion Laboratory, California Institute of Technology,
under a contract with the National Aeronautics and Space
Administration.  The National Radio Astronomy Observatory is a
facility of the National Science Foundation operated under cooperative
agreement by Associated Universities, Inc.  LB and JM acknowledge
support from Chilean Center of Excellence in 
Astrophysics and Associated Technologies (PFB 06), and from Chilean
Center for Astrophysics FONDAP 15010003.  This research made use of
Montage, funded by the National Aeronautics and Space Administration's
Earth Science Technology Office, Computation Technologies Project,
under Cooperative Agreement Number NCC5-626 between NASA and the
California Institute of Technology. Montage is maintained by the
NASA/IPAC Infrared Science Archive.

\begin{table*}
\centering
\space
\caption{CBI Fields}
\label{tbl:fields}
\begin{tabular}{|c||c|c|c|c|}
\hline\hline
& & & &  \\
name & R.A. & Dec & Dimensions & Ground Removal \\
     & (J2000) & (J2000) &     & Strategy \\
 & & & & \\
\hline
 & & & & \\
02h & $02:50:00$ & $-01:30:00$ & $5^{\circ} \times 6^{\circ}$ & 8-min differences \\
08h-deep & $08:44:40$ & $-03:10:00$ & $0.75^{\circ} \times 0.75^{\circ}$& 8-min differences\\
14h & $14:50:00$ & $-02:30:00$ & $5^{\circ} \times 5^{\circ}$ & 8-min differences\\
14h-deep & $14:42:00$ & $-03:50:00$ & $0.75^{\circ} \times 0.75^{\circ}$& 8-min differences\\
20h & $14:50:00$ & $-02:30:00$ & $5^{\circ} \times 5^{\circ}$ & 8-min differences\\
20h-deep & $20:48:40$ & $-03:30:00$ & $0.75^{\circ} \times 0.75^{\circ}$& 8-min differences\\
02h-pol & $02:49:30$ & $-02:52:30$ & $5^{\circ} \times 5^{\circ}$ & scan mean subtraction \\
08h-pol & $08:47:30$ & $-02:47:30$ & $5^{\circ} \times 5^{\circ}$ & scan mean subtraction \\
14h-pol & $14:45:30$ & $-04:07:30$ & $5^{\circ} \times 5^{\circ}$ & scan mean subtraction \\
20h-pol-deep & $20:49:30$ & $-03:30:00$ & $5^{\circ} \times 0.75^{\circ}$ & scan mean subtraction \\
\hline\hline
\end{tabular}
\end{table*}

\begin{table*}
\centering
\space
\caption{CBI Power Spectrum.  Power spectrum from the total CBI dataset.  Columns are 1)
  bin, 2) power spectrum, 3) Gaussian error on the power spectrum, 4)
  thermal noise power in the bin, 5) the contribution from unresolved
  source in the bin (assuming 0.046 $\jysq$), 6) lower $\ell$ limit, and 7) upper
  $\ell$ limit.}
\label{tbl:powspec}
\begin{tabular}{|c||c|c|c|c|c|c|}
\hline \hline
$\ell$ & $C_\ell (\mu \rm{K}^2)$ & Error & $C_{noise} (\mu\rm{K}^2)$ & $C_{src}
(\mu\rm{K}^2) $ & $\ell_{min}$
& $\ell_{max}$  \\
\hline

  1 & 5695.372 &  753.175 & 1767.465 &   2.4 &   0.00 &   350.00 \\
  2 & 1260.879 &  260.277 &  218.186 &   1.2 &  350.00 &   450.00 \\
  3 & 2987.370 &  408.622 &  300.134 &   3.5 &  450.00 &   540.00 \\
  4 & 2449.122 &  376.215 &  235.321 &   4.7 &  540.00 &   620.00 \\
  5 & 1844.757 &  268.193 &  311.309 &   3.5 &  620.00 &   700.00 \\
  6 & 2680.066 &  346.136 &  487.124 &   4.7 &  700.00 &   780.00 \\
  7 & 2241.459 &  312.705 &  464.681 &   6.0 &  780.00 &   860.00 \\
  8 & 2204.210 &  313.916 &  602.316 &   8.3 &  860.00 &   940.00 \\
  9 &  691.900 &  221.538 &  600.919 &   9.9 &  940.00 &  1020.00 \\
 10 &  979.649 &  220.328 &  555.848 &   9.5 & 1020.00 &  1100.00 \\
 11 & 1166.825 &  203.100 &  607.624 &  11.9 & 1100.00 &  1192.00 \\
 12 & 1041.110 &  166.131 &  669.178 &  14.2 & 1192.00 &  1302.00 \\
 13 &  727.473 &  143.316 &  706.893 &  13.7 & 1302.00 &  1425.00 \\
 14 &  803.088 &  147.134 & 1218.974 &  19.7 & 1425.00 &  1600.00 \\
 15 &  311.402 &  126.647 & 1240.392 &  27.4 & 1600.00 &  1800.00 \\
 16 &  379.102 &  122.922 & 1316.752 &  29.9 & 1800.00 &  2050.00 \\
 17 &  260.743 &  132.234 & 1666.893 &  35.4 & 2050.00 &  2350.00 \\
 18 &  386.738 &  117.427 & 3000.407 &  66.1 & 2350.00 &  3900.00 \\
\hline \hline
\end{tabular}
\end{table*}

\begin{table*}
\centering
\space
\caption{Cosmological Parameters from CBI+WMAP5.  These are the standard basic 6 parameters of the tilted
  $\Lambda$CDM model and the SZ amplitude $q_{SZ}$, plus parameters
  derived from them for CBI and WMAP5.  Note that all of the parameters except for
  $\sigma_8$, and the $\sigma_8$ inferred from the SZ amplitude
  $\sigma_8^{SZ} \propto q_{SZ}^{1/7}$  are relatively insensitive to
  the two SZ templates used.  All parameters above the line are
  independent variables, and those below are derived from the
  independent parameters.}  

\label{tbl:wmap5cbiparams}
\begin{tabular}{|c||c||c|}
\hline\hline
Parameter & CBI+WMAP5+KS sz & CBI+WMAP5+Bond sz  \\
\hline
$    \Omega_b h^2 $  &  $ 0.02291 \pm 0.00061 $  &  $ 0.02271 \pm
0.00060 $  \\
$    \Omega_c h^2 $  &  $ 0.1069 \pm 0.0064 $  &  $ 0.1081 \pm 0.0063
$  \\
$    \theta $  &  $ 1.0406 \pm 0.0030 $  &  $ 1.0404 \pm 0.0030 $  \\
$    \tau $  &  $ 0.087 \pm 0.018 $  &  $ 0.086 \pm 0.018 $  \\
$    q_{SZ} $  &  $ 2.52 \pm 0.96 $  &  $ 5.3 \pm 1.8 $  \\
$    n_s $  &  $ 0.960 \pm 0.015 $  &  $ 0.963 \pm 0.015 $  \\
$    log[10^{10} A_s] $  &  $ 3.039 \pm 0.044 $  &  $ 3.052 \pm 0.042
$  \\
\hline
$    \Omega_\Lambda $  &  $ 0.756 \pm 0.030 $  &  $ 0.750 \pm 0.030 $
\\
$    Age/GYr $  &  $ 13.65 \pm 0.14 $  &  $ 13.68 \pm 0.14 $  \\
$    \Omega_m $  &  $ 0.244 \pm 0.030 $  &  $ 0.250 \pm 0.030 $  \\
$    \sigma_8 $  &  $ 0.770 \pm 0.038 $  &  $ 0.781 \pm 0.036 $  \\
$    z_{re} $  &  $ 10.8 \pm 1.5 $  &  $ 10.9 \pm 1.5 $  \\
$    \sigma_8^{SZ} $  &  $ 0.910 \pm 0.064 $  &  $ 1.015 \pm 0.060
$  \\
$    H_0 $  &  $ 73.2 \pm 2.8 $  &  $ 72.6 \pm 2.8 $  \\
\hline\hline
\end{tabular}
\end{table*}
 
\begin{table*}
\centering
\space
\caption{Cosmological Parameters Determined from the Combined CMB
  Datasets.  These are the standard basic 6 parameters of the tilted
  $\Lambda$CDM model and the SZ amplitude $q_{SZ}$, plus parameters
  derived from them for combined CMB datasets (see text for the list).
  Columns are as in Table~\ref{tbl:wmap5cbiparams}. 
}  
\label{tbl:wmap5cmballparams}
\begin{tabular}{|c||c||c|}
\hline\hline
Parameter & WMAP5+CMBall+KS SZ & WMAP5+CMBall+SPH SZ  \\
\hline
$    \Omega_b h^2 $  &  $ 0.02289 \pm 0.00057 $  &  $ 0.02264 \pm
0.00056 $  \\
$    \Omega_c h^2 $  &  $ 0.1073 \pm 0.0068 $  &  $ 0.1088 \pm 0.0054
$  \\
$    \theta $  &  $ 1.0419 \pm 0.0025 $  &  $ 1.0417 \pm 0.0025 $  \\
$    \tau $  &  $ 0.085 \pm 0.018 $  &  $ 0.085 \pm 0.017 $  \\
$    \alpha_{SZ} $  &  $ 2.65 \pm 0.81 $  &  $ 4.3 \pm 1.3 $  \\
$    n_s $  &  $ 0.956 \pm 0.014 $  &  $ 0.960 \pm 0.014 $  \\
$    log[10^{10} A_s] $  &  $ 3.033 \pm 0.039 $  &  $ 3.051 \pm 0.038
$  \\
\hline
$    \Omega_\Lambda $  &  $ 0.758 \pm 0.026 $  &  $ 0.749 \pm 0.027 $
\\
$    Age/GYr $  &  $ 13.62 \pm 0.13 $  &  $ 13.66 \pm 0.13 $  \\
$    \Omega_m $  &  $ 0.242 \pm 0.026 $  &  $ 0.251 \pm 0.027 $  \\
$    \sigma_8 $  &  $ 0.769 \pm 0.031 $  &  $ 0.784 \pm 0.030 $  \\
$    z_{re} $  &  $ 10.7 \pm 1.5 $  &  $ 10.8 \pm 1.5 $  \\
$    \sigma_8^{SZ} $  &  $ 0.922 \pm 0.047 $  &  $ 0.988 \pm 0.049
$  \\
$    H_0 $  &  $ 73.5 \pm 2.6 $  &  $ 72.6 \pm 2.6 $  \\
\hline\hline
\end{tabular}
\end{table*}

\begin{table*}
\centering
\space
\caption{CBI Predictions for Other Experiments.  For convenience in comparing the CBI results with other
  experiments if the excess power is due to SZ, we present the signals
  predicted by the two SZ templates at a variety of frequencies and
  angular scales.  All values are in $\mu \rm{K}^2$.}
\label{tbl:excesspred}
\begin{tabular}{|c||c||c||c||c||c||c|}
\hline\hline
$\ell$ & 30 GHz KS & 100 GHz KS & 150 GHz KS & 30 GHz SPH & 100 GHz SPH & 150 GHz SPH  \\
\hline

 500   &  48.6 $\pm$  18.5  &  29.0 $\pm$  11.0  &  11.6 $\pm$   4.4  &  24.3 $\pm$   8.3  &  14.5 $\pm$   4.9  &   5.8 $\pm$   2.0 \\
1000   &  73.3 $\pm$  27.9  &  43.6 $\pm$  16.6  &  17.4 $\pm$   6.6  &  52.5 $\pm$  17.8  &  31.3 $\pm$  10.6  &  12.5 $\pm$   4.2 \\
1500   &  87.8 $\pm$  33.4  &  52.3 $\pm$  19.9  &  20.9 $\pm$   8.0  &  94.2 $\pm$  32.0  &  56.1 $\pm$  19.1  &  22.4 $\pm$   7.6 \\
2000   &  94.2 $\pm$  35.9  &  56.1 $\pm$  21.4  &  22.4 $\pm$   8.5  & 138.2 $\pm$  46.9  &  82.3 $\pm$  28.0  &  32.9 $\pm$  11.2 \\
2500   &  96.2 $\pm$  36.6  &  57.3 $\pm$  21.8  &  22.9 $\pm$   8.7  & 151.8 $\pm$  51.5  &  90.4 $\pm$  30.7  &  36.1 $\pm$  12.3 \\
3000   &  96.2 $\pm$  36.7  &  57.3 $\pm$  21.8  &  22.9 $\pm$   8.7  & 189.3 $\pm$  64.3  & 112.8 $\pm$  38.3  &  45.0 $\pm$  15.3 \\
4000   &  92.4 $\pm$  35.2  &  55.1 $\pm$  21.0  &  22.0 $\pm$   8.4  & 232.9 $\pm$  79.1  & 138.8 $\pm$  47.1  &  55.4 $\pm$  18.8 \\
5000   &  86.2 $\pm$  32.8  &  51.4 $\pm$  19.6  &  20.5 $\pm$   7.8  & 280.4 $\pm$  95.2  & 167.0 $\pm$  56.7  &  66.7 $\pm$  22.7 \\
6000   &  80.3 $\pm$  30.6  &  47.9 $\pm$  18.2  &  19.1 $\pm$   7.3  & 294.0 $\pm$  99.9  & 175.2 $\pm$  59.5  &  70.0 $\pm$  23.8 \\
8000   &  67.4 $\pm$  25.7  &  40.1 $\pm$  15.3  &  16.0 $\pm$   6.1  & 316.9 $\pm$ 107.6  & 188.8 $\pm$  64.1  &  75.4 $\pm$  25.6 \\

\hline\hline
\end{tabular}
\end{table*}

\begin{figure*}[!t]
\centering
\includegraphics[width=0.95\textwidth]{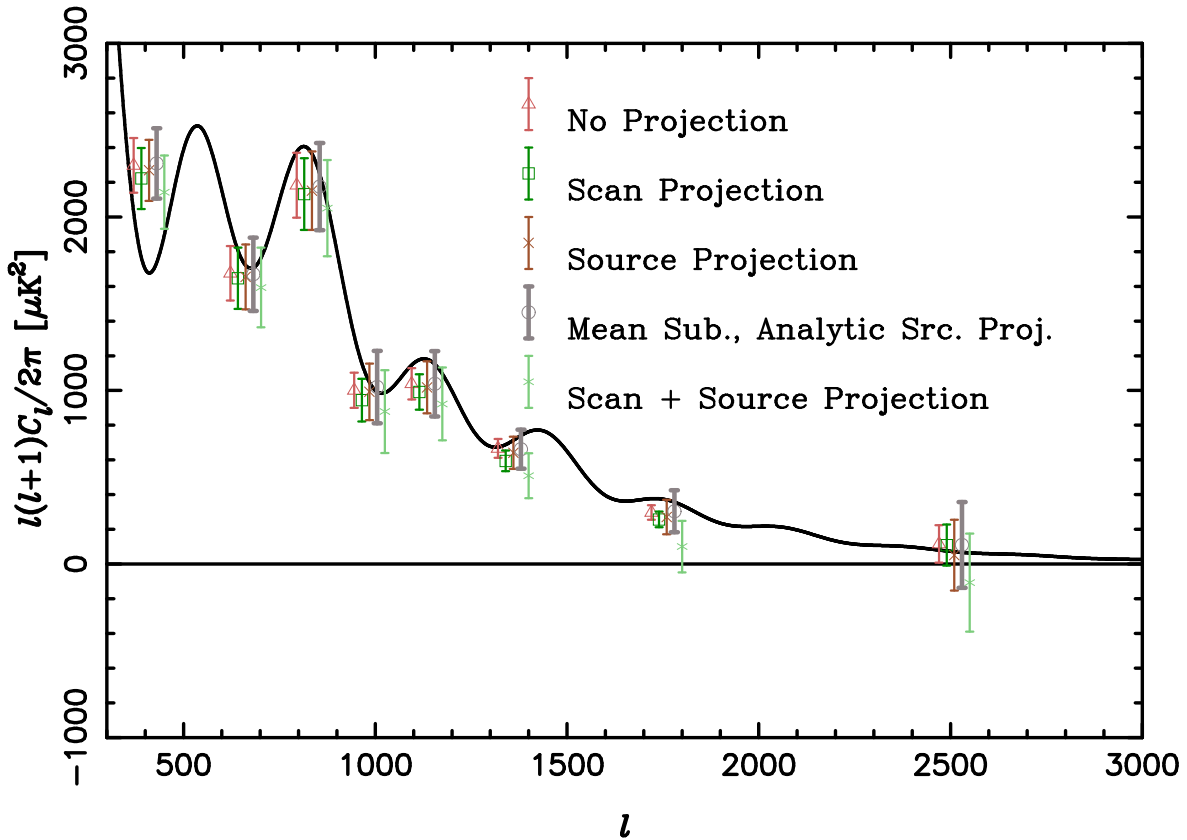}
\caption{Effects of projections on the expected spectrum from the CBI
  strips.  The red triangles show the expected spectrum from just the
  TT part of the CBI polarization data with no removal of ground or
  sources.  The green squares show the expected spectrum when the scan
  mean (which contains the ground signal) is projected out.  The red
  X's show the spectrum when only the sources are projected.  The
  green X's show the spectrum when both sources and ground are
  projected - there is a significant bias downwards in the spectrum.
  The grey circles show the the same spectrum when the scan mean is
  subtracted rather than projected, and the sources are projected
  using the numerically stable method of Appendix \ref{sec:maxlike}.
  The black curve is the input CMB spectrum.
  This method keeps the covariance matrix better conditioned,
  so it is less susceptible to roundoff errors in the inversion.
  These spectra are fit to the theoretical CMB+noise signal matrix
  using the techniques described in Appendix \ref{sec:spec_expected},
  which is equivalent to averaging over all possible noise and
  signal simulations.  The CBI differenced data from \citet{paper7}
  were already effectively scan-mean subtracted, and so are not
  included here.}
\label{fig:Bias}
\end{figure*}

\bibliographystyle{apj}
\bibliography{ms.bib}

\appendix

\section{Maximum Likelihood Fitting}
\label{sec:maxlike}

We use the program MPILIKELY to measure the maximum-likelihood power
spectrum and related quantities from the gridded data and its noise
and signal matrices.  MPILIKELY is an MPI implementation of the
algorithm described in \citet{sieversthesis}.  We briefly summarize
the algorithm here, as well as describe additional features of
MPILIKELY.  

\subsection{Fast Curvature and Gradient Calculation}
\label{sec:curvederiv}

The likelihood of correlated Gaussian data is:
\begin{equation}\label{eq:Likelihood}
  \log \left ( \mathcal{L} \right ) = -\frac{1}{2}\data^\dag\cov^{-1}\data -
  \frac{1}{2}\log \left ( | \cov | \right )
\end{equation}
where $\data$ is the data, and $\cov =\left < \data_i \data_j \right >
$ is the covariance matrix( \eg\  \citet{BJK98}).  The correlation matrix in general will depend on
both the noise and the signal in the data.  The maximum-likelihood
solution is then the set of parameters on which $\cov$ depends that
maximizes the likelihood.  It is often the case that $\cov$ depends
linearly on its parameters.  This is true if we
parameterize the CMB power spectrum by bands in $\ell$, in which case the theory
matrix of Equation \ref{eq:ccovar} takes the form
\begin{equation}\label{eq:covmat}
\cov_T=\sum q_B \cov_B,
\end{equation}
in terms of CMB signal matrices $\cov_B$ with associated bandpowers
$q_B$.  The standard technique for
maximizing the likelihood is to calculate the gradient and curvature
of the log-likelihood and take a multi-dimensional Newton's method step,
iterating until convergence.  The gradient of the log likelihood when
the theory covariance is of the form given in Equation \ref{eq:covmat} is:
\begin{equation}
\frac{\partial \log \left ( \mathcal{L} \right )}{\partial q_B} =
\frac{1}{2} \data^\dag \cov^{-1} \cov_B \cov^{-1}\data - \frac{1}{2} \rm{Tr} \left( \cov^{-1}\cov_B \right )
\end{equation}

where $Tr$ is the trace operator. The curvature is:

\begin{equation}
\frac{\partial^2 \log  \mathcal{L}}{\partial q_B  \partial q_B'} 
= -\data^\dag \cov^{-1}\cov_B \cov^{-1} \cov_{B'} \cov^{-1}\data + \frac{1}{2}
  Tr \left ( \cov^{-1}\cov_B \cov^{-1}\cov_{B'} \right )
\end{equation}

.  It is straightforward to show that 
\begin{equation}\label{eq:chisq_trace}
 \bvec{a}^\dag {\bf B}\bvec{a} = \rm{Tr} \left ( {\bf B}  \bvec{aa}^\dag \right ) \end{equation}
for vector $\bvec{a}$ and matrix ${\bf B}$.  One can use that identity
to rewrite the first term in the curvature as follows:
\begin{equation} \label{eq:curve_term1}
\rm{Tr} \left [ \cov^{-1} \cov_B \cov^{-1} \cov_{B'} \cov^{-1} \data\data^T \right ].
\end{equation}
If the correlation matrix is a good description of the data, we
have $\left < \data \data^T \right > = \cov$, or equivalently that
$\cov^{-1} \data \data ^T
\simeq {\bf I}I$, the identity matrix.  At this point, the standard treatment
\citep[\eg]{Bond98} is to replace $\cov^{-1}\data\data^T$ with ${\bf I}$ which leaves 
\begin{equation}\label{eq:slowcurve}
\frac{\partial^2 \log \left ( \mathcal{L} \right )}{\partial q_B \partial q_B'} 
\simeq  - \frac{1}{2} \rm{Tr} \left ( \cov^{-1}\cov_B \cov^{-1}\cov_{B'}
\right ) .
\end{equation}
The most efficient implementation when using this approximation
requires one to pre-calculate the set matrices $\cov^{-1}\cov_B$,
which requires an expensive matrix-matrix multiplication for each
bandpower.  One can then use the fact that $\rm{Tr} \left ( {\bf AB} \right )$ is
an $n^2$ operation (essentially since one only need calculate the
diagonal elements of ${\bf AB}$).  

Instead of using $\left < \data \data^T \right > \approx
\cov$ to cancel the first term in the curvature, MPILIKELY uses it to
cancel the second term, leaving:
\begin{equation}\label{eq:fastcurve}
\frac{\partial^2 \log \left ( \mathcal{L} \right )}{\partial q_B
  \partial q_B'} \simeq -\frac{1}{2}\data^T \cov^{-1}\cov_B \cov^{-1}
  \cov_{B'} \cov^{-1}\data. 
\end{equation}
The great advantage of this form of the approximate curvature is
that it can be calculated using strictly matrix-vector operations.  In
practice, on MPI clusters, we find that MPILIKELY typically spends
80-90\% of its time inverting the covariance matrix, even when fitting
dozens of parameters.  In addition, storage requirements are halved
since we don't have to store the set of matrices $\cov^{-1}\cov_B$.
We also note that the error in Equation 
\ref{eq:slowcurve} is exactly twice that of \ref{eq:fastcurve}.
However, since the curvature only affects the path taken to the
maximum likelihood solution, both approximations will result in the
same bandpowers.  For final data products, we typically evaluate the
full curvature at convergence for more accurate error bars.

When finding the maximum-likelihood spectrum, MPILIKELY uses a
modified version of the Levenberg-Marquardt algorithm (see \eg\
\citet{NumRec}) that generally reduces to Newton-Raphson iteration.
The Levenberg-Marquardt control parameter $\lambda$ is initially set
to zero, and remains there as long as the covariance remains
positive-definite.  On the first failure, it is set to unity, under
the theory that if the current guess has overshot the maximum enough
to make $\cov$ non-positive-definite, a correction of the step by of
order at least a factor of two is warranted.  On continued failures,
we increase $\lambda$ by a factor of 2, and on successes, we decrease
it by $\sqrt{2}$ until $\lambda$ falls below one, at which point we
set it to zero.  Our convergence criteria are that the largest step in
any dimension, in terms of the error in that dimension, is less than
some small fraction ( typically 0.01), and that $\lambda=0$ during
that iteration.  In practice, $\lambda$ stays at zero, unless there is
a power spectrum bin that is significantly negative.  This happens
when a random realization of the noise has substantially less power in
it than expected, which pushes the power spectrum negative and
introduces substantial skewness to the likelihood.

\subsection{Spectrum From Matrices}
\label{sec:spec_expected}

It is useful to be able to calculate the power spectrum and errors
expected from either a noiseless data vector or a signal matrix.  The
treatment for the two cases is similar.  We can invoke the identity in
Equation \ref{eq:chisq_trace}, replace $\data\data^T$ with the matrix
${\bf D}$, and
take advantage of the fact that $Tr\left ( {\bf AB} \right ) = Tr
\left ( {\bf BA}\right ) $ to rewrite the gradient as follows:
\begin{equation}
\frac{\partial \log \left ( \mathcal{L} \right )}{\partial q_B} = 
\frac{1}{2} \rm{Tr} \left ( \cov^{-1} {\bf D} \cov^{-1} \cov_B  \right ) -
\frac{1}{2} \rm{Tr} \left( \cov^{-1}\cov_B \right ) .
\end{equation}
We can now find the expected spectrum from data drawn from
covariance matrix ${\bf D}$, marginalized over realizations of the data.
When fitting to noiseless data, we have ${\bf D} = \data\data^T + \cov^{N}$ to marginalize
over realizations of the noise.  The gradient calculation is slowed by
a factor of $\sim 3$ since we now have to calculate $\cov^{-1}{\bf D}\cov^{-1}$
instead of just $\cov^{-1}$.  The curvature calculation is slightly more
complicated.  Without an actual data vector, we cannot take advantage
of the fast curvature calculation in Equation \ref{eq:fastcurve}.
While one could use the curvature in Equation \ref{eq:slowcurve}, this
can become prohibitively slow in practice.  MPILIKELY's solution is to
draw a set of sample data vectors from ${\bf D}$, and then average Equation
\ref{eq:fastcurve} for each of those realizations.  The only $n^3$
operation this requires is a single initial Cholesky factorization of
${\bf D}$ to calculate the data vectors.  

\subsection{Source Projection}
\label{sec:sourceproj}

The standard way of removing point sources with known positions is by
projecting them from the covariance matrix \citep[\eg\
]{Bond98,paper2}.  This is roughly equivalent to masking out the source
location in the map.  In practice, this is done by adding $\csrc = \beta s_is_i^\dag$ to
the covariance matrix, where $s_i$ is the signal expected from the
$i^{th}$ source, and $\beta$ is an (extremely large) amplitude.  For
sufficiently large $\beta$, the power spectrum is insensitive to any
signal proportional to $s_i$ - \ie\ we don't need to know the
amplitude of the source.  We have previously used large values of
$\beta$, the projection amplitude, but this can lead to numerical
stability problems as large $\beta$'s cause $\cov$ to become
ill-conditioned.  Instead, in MPILIKELY we take the analytic limit as
$\beta \rightarrow \infty$ using the Woodbury formula \citep[see \eg\
]{NumRec}.  For symmetric matrices, we have
\begin{equation}\label{eq:sherwood}
\lim_{\beta\to\infty} 
\left ( \cov  + \beta {\bf SS}^\dag\right )^{-1} =
 \cov^{-1} -
\cov^{-1} {\bf S} \left ( {\bf S}^\dag
\cov^{-1} {\bf S} \right )^{-1} {\bf S}^\dag \cov^{-1}.
\end{equation}
One must also guard against the possibility of ${\bf S}$ being
degenerate.  If it is, the calculation of $\left ( {\bf S}^\dag
\cov^{-1} {\bf S}\right )^{-1}$ will fail.  To deal with this
possibility, we first scan ${\bf S}$ for repeated columns (same source
entered twice).  We then orthogonalize ${\bf S}$ through use of a QR
factorization, and use the orthogonal matrix ${\bf Q}$ in Equation
\ref{eq:sherwood} instead of ${\bf S}$.  One could also use SVD, but
QR is typically a factor of $\sim 6$ faster.  With the addition of the
QR factorization, the Woodbury formula never requires the inversion of
an ill-conditioned matrix, and so as long as the source-free version
of the covariance $\cov$ is well-conditioned, source projection does
not lead to numerical instability in calculating the source-projected $\cov^{-1}$.

\subsection{Likelihood Evaluations}
\label{sec:likelihood}
It is often useful to directly evaluate the likelihood of a given
covariance matrix.  Some instances where this is useful are when
comparing the goodness-of-fit of two different models for the data, or
when measuring quantities like confidence intervals that depend on the
non-Gaussian nature of the likelihood surface.  The analytic
projection of sources using the Woodbury formula complicates
likelihood evaluations because $\cov^{-1}$ becomes singular (the
projected modes are truly gone).  The solution is to rotate into the
subspace spanned by the complement of the source vectors.  The
rotation/compression matrix can also be found efficiently through a QR
factorization of the source vectors (most QR implementations also have
a way to find an orthogonalized matrix spanning the complement of the
original matrix).  Each matrix (noise, banded signal...) and the data
are then compressed.  One could fit the spectrum with these compressed
matrices, which gives the same spectrum one gets when using Equation
\ref{eq:sherwood}.  Since the matrices are smaller, the actual fitting
of the power spectrum is sped up.  However, the initial overhead
required (2 matrix-matrix multiplications for each CMB band, plus
typically a few others) is generally much larger than the total time
to fit the spectrum with the uncompressed matrices, and so we in
general only compress the matrices if we desire likelihoods.

\subsection{Template Fitting}
\label{sec:templates}

MPILIKELY also supports simultaneous fitting of additive templates to
the data when measuring the power spectrum.  We use this for measuring
the amplitude of foreground maps in the CBI data.  For additive
templates, the template model is subtracted from the data before
calculating the likelihood, which then becomes:

\begin{equation}\label{eq:template_likelihood}
  \log \left ( \mathcal{L} \right ) = -\frac{1}{2}\data^{*\dag}\cov^{-1}\data^{*} -
  \frac{1}{2}\log \left ( | \cov | \right )
\end{equation} 
where $\data^* = \data-\sum a_j \bvec{m}^j$ for expected signal
$\bvec{m}^j$ with amplitude $a_j$ and observed data $\data$.  Here,
the index $j$ runs over templates, not over data elements.  The
gradient and curvature of the CMB terms are unchanged, as long as they
are calculated using $\data^*$ instead of $\data$.  The template terms
in the gradient are:

\begin{equation}
\frac{\partial \log \left ( \mathcal{L} \right )}{\partial a_j} =
\bvec{m}^{j \dag} \cov^{-1} \data^* .
\end{equation}
The curvature terms are:

\begin{equation}
\frac{\partial^2 \log \left ( \mathcal{L} \right )}{\partial a_j
  \partial a_{j'}} =
-\bvec{m}^{j \dag} \cov^{-1} \bvec{m}^{j'}
\end{equation}
and

\begin{equation}
\frac{\partial^2 \log \left ( \mathcal{L} \right )}{\partial a_j
  \partial q_B} =
-\bvec{m}^{j \dag} \cov^{-1}\cov_B \cov^{-1} \data .
\end{equation}
These curvature and gradient terms only require matrix-vector
operations to calculate, and so have a negligible impact on the time
required to fit the power spectrum.

\end{document}